\documentclass[sn-standardnature]{sn-jnl}
\usepackage{xcolor}
\usepackage{soul}

\jyear{2022}%

\theoremstyle{thmstyleone}%
\theoremstyle{thmstyletwo}%

\theoremstyle{thmstylethree}%
\begin{document}

\title[ ]{An Origami-Inspired Design of Highly Efficient Cellular Cushion Materials
}


\author[1]{\fnm{Ahmed} \sur{S. Dalaq}}\email{asd9759@nyu.edu}

\author[1,2]{\fnm{Shadi} \sur{Khazaaleh}}\email{smk24@nyu.edu}

\author*[1,2]{\fnm{Mohammed} \sur{F. Daqaq}}\email{mfd6@nyu.edu}

\affil[1]{\orgdiv{Engineering Division}, \orgname{New York University Abu Dhabi (NYUAD)}, \orgaddress{ \city{Saadiyat Island},  \country{UAE}}}

\affil[2]{\orgdiv{Department of Mechanical and Aerospace Engineering}, \orgname{Tandon School of Engineering, NYU}, \city{Brooklyn}, \postcode{11201}, \state{NY}, \country{USA}}


\abstract{\raggedright Current architectured cellular cushion materials rely mainly on damage and/or unpredictable collapse of their unit cells to absorb and dissipate energy under impact. This prevents shape recovery and produces undesirable force fluctuations that limit reusability and reduce energy absorption efficiency. Here, we propose to combine advanced manufacturing technologies with Origami principles to create a new class of architectured cellular viscoelastic cushion material which combines low weight and high energy absorption efficiency with damage resistance and full behavior customization. Each unit cell in the proposed material is inspired by the Kresling Origami topology, which absorbs impact energy by gracefully folding the different interfaces forming the cell to create axial and rotational motions. A large part of the absorbed energy is then dissipated through viscoelasticity and friction between the interfaces. The result is a nearly ideal cushion material exhibiting high energy absorbing efficiency ($\sim$ 70\%) combined with high energy dissipation (94\% of the absorbed energy). The material is also tunable for optimal performance, reliable despite successive impact events, and achieves full shape recovery. }




\maketitle

\section{Introduction}\label{sec1}
\begin{figure}[t]%
\centering
\includegraphics[width=\textwidth]{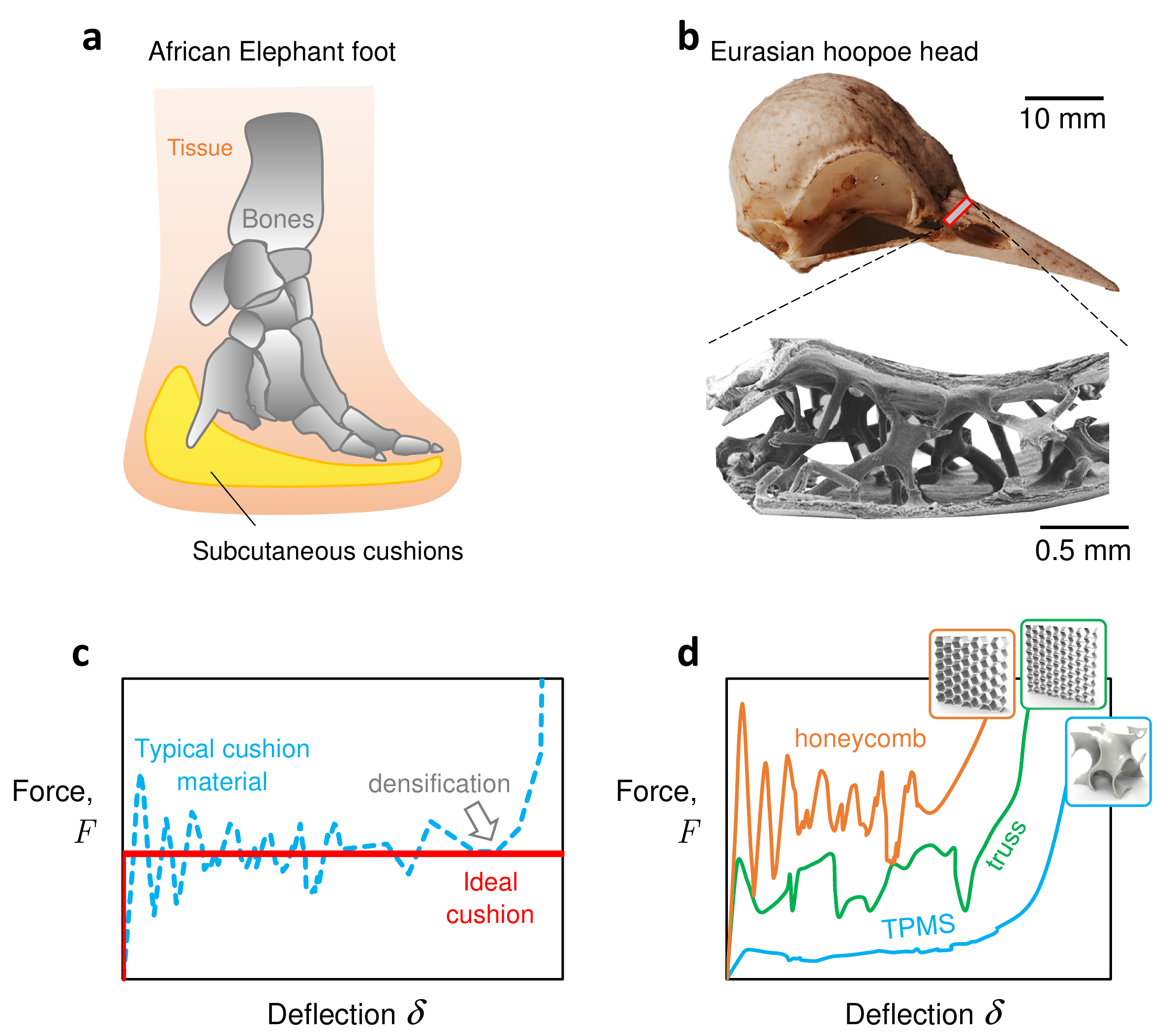}
\caption{ \textbf{a}  Foot anatomy of the African Elephant (\textit{Loxodonta Africana} \cite{weissengruber_structure_2006}). \textbf{b} Skull of the Eurasian Hoopoe (\textit{Upupa Epops} \cite{wang_why_2011}). \textbf{c}  $F-\delta$ curve of an ideal cushion material (in solid red) and a typical one (in dashed blue). Onset of densification is indicated by a grey arrow. \textbf{d} $F-\delta$ curves of some common cellular architectures: honeycomb \cite{kucewicz_modelling_2018}, truss \cite{mohsenizadeh_additively-manufactured_2018}, and triply periodic minimal sheets (TPMS) \cite{al-ketan_nature-inspired_2018}.}
\label{fig1}
\end{figure}

From the fins of the humpback whale (\textit{megaptera novaeangliae} \cite{clapham_megaptera_1999}) inspiring the design of quieter stall-resistant turbine blades to the leaves of the lotus flower (\textit{nelumbo nucifera} \cite{feng_super-hydrophobic_2002}) providing the fundamental knowledge necessary to create superhydrophobic surfaces, nature continues to inspire many of today's most efficient manmade inventions. And while nature did not directly inspire the design of traditional cushion materials, it provided hints regarding the elements necessary to build efficient ones \cite{ha_review_2020}. In particular, animals that are constantly subjected to high impact forces have evolved to incorporate porous and/or viscoelastic elements. Porosity serves to absorb kinetic energy through compaction, while viscosity serves to dissipate a part of the absorbed energy as heat. An interesting illustration lies in the feet of the African elephants (\textit{loxodonta africana} \cite{weissengruber_structure_2006}), which has a subcutaneous cushion consisting of layers of fatty viscoelastic fibrous connective tissue to dissipate large stresses induced by the elephant's weight during locomotion (see Fig. \ref{fig1}a). Another intriguing example can be seen in the eurasian hoopoe’s (\textit{upupa epops} \cite{wang_why_2011}) porous cranial bone, which protects its brain against high-frequency impacts reaching up to an acceleration of 1000 g (see Fig. \ref{fig1}b). 

Interestingly, current engineered cushion materials, which are mostly polymeric foams do, to some extent, mimic those optimized by nature through a lengthy process of natural selection and evolution. They are disordered cellular viscoelastic structures with open or closed cells that have high crushability and energy absorption capacity \cite{gibson_mechanics_1982,aubert_low-density_1985,brumfield_characterization_1969, almanza_microestructure_2001}. Their performance, and that of other energy absorbing materials, is a function of their internal structure and density, and is assessed based on the shape of their force displacement ($F-\delta$) curve  during quasi-static loading or impact \cite{gibson_cellular_1999,gibson_mechanics_1982}. Besides having a large area under its $F-\delta$ curve, which represents the amount of raw energy absorbed by the material during loading, a good cushion material should exhibit a sharp increase in force upon impact with minimal force overshoot to avoid damaging protected subjects \cite{xiang_energy_2020, wang_architected_2019} (Fig. \ref{fig1}c). Upon reaching the maximum allowable force level, the material should maintain a constant resistance to displacement/penetration without force fluctuation. In addition, unlike the blue dashed curve in Fig. \ref{fig1}c, an ideal cushion material should not experience full densification prior to full energy dissipation. It should be damage resistant and able to sustain multiple impacts with minimal deterioration in its effectiveness; or better yet, the material should be able to recover its initial shape after impact. Finally, the ideal material should be lightweight, low cost, and easy to install.

To realize the aforedescribed properties of the ideal cushion material, much of the earlier research work focused on manipulating the density of the material and the average size of the voids to offer some control over the maximum forces experienced by the subject and to maximize energy absorption with minimal densification \cite{ashby_mechanical_1983,lakes_materials_1993,gosselin_cell_2005, gibson_biomechanics_2005, schaedler_ultralight_2011}. More recently, as a result of the advent of 3D/4D printing \cite{kuang_advances_2019, thakar_3d_2022, teunis_4d_2021}, new avenues were opened towards the design of versatile architectured cellular cushion materials that can combine high strength and toughness \cite{dalaq_finite_2016, han_microscopic_2017, jiang_highly-stretchable_2016,dalaq_mechanical_2016, abou-ali_mechanical_2020}, high stiffness-to-weight ratio \cite{habib_fabrication_2018, li_additive_2021-1, abueidda_effective_2016}, and compactness with other desirable thermal transport properties \cite{qureshi_effect_2022, abueidda_micromechanical_2015, baobaid_fluid_2022, alqahtani_thermal_2021}. Such materials, which are formed of unit cells of different geometric features arranged in various configurations, seem ideal for the design of highly-efficient energy absorbing cushions with tunable $F-\delta$ curves.

Yet, despite the introduction of architectured cellular materials, many of the ideal cushion qualities, such as damage resistance, shape recovery, and absence of overshoots and fluctuations are still challenging to fulfill. This is because much of the current cellular materials utilize damage as an energy dissipation mechanism \cite{dalaq_mechanical_2016, habib_fabrication_2018}, or they rely on sudden collapse and unstable buckling of their features \cite{abueidda_compression_2020, song_octet-truss_2019, gibson_mechanics_1982}, all of which are unpredictable or stochastic at best. Take for example, Fig. \ref{fig1}d, which demonstrates the $F-\delta$ curves for three engineered cellular cushion materials. Both of the honeycomb and truss architectures exhibit large force overshoots and non-smooth deformations under stresses \cite{kucewicz_modelling_2018, mohsenizadeh_additively-manufactured_2018}. Triply periodic minimal sheets (TPMS), on the other hand, have a much smoother $F-\delta$ curve, but still rely mainly on damage to absorb energy and show limited recovery after impact \cite{abueidda_compression_2020, dalaq_mechanical_2016, abou-ali_mechanical_2020}. 

Creating a reliable $F-\delta$ curve that fulfills the qualities of an ideal cushion remains a challenge in the formation of modern cushion materials. So much so that some investigators have utilized pneumatic systems in cellular materials \cite{wang_architected_2019}. To circumvent this critical problem, we propose in this work to combine 3D printing of viscoelastic materials \cite{lakes_viscoelastic_1998, ramirez_viscoelastic_2018} with Origami principles \cite{lang_science_2007, kresling_fifth_2020} to create a new class of cushion materials that is tunable, recoverable, and has a predictable response. The choice of Origami-inspired unit cells to construct the proposed cellular cushion is not arbitrary, but rather motivated by the unique fundamental mechanics of a special Origami topology called the Kresling pattern. In particular, the special geometry of the Kresling unit cell permits graceful folding at different interfaces under loading, which, in turn, forms large recoverable axial and rotational deformations that are free of permanent damage or irreversible buckling \cite{dalaq_experimentally-validated_2022,kresling_origami-structures_2012, kresling_fifth_2020, khazaaleh_combining_2022, zhai_origami-inspired_2018}. This, combined with the fact that the geometric properties of each unit cell has a qualitative and predictable influence on the restoring force behavior of the material, permits customizing the $F-\delta$ curve for the application under consideration. This includes designing linear, nonlinear softening, hardening, or even bi-stable cushions \cite{dalaq_experimentally-validated_2022, khazaaleh_combining_2022}. Finally, the process of folding and unfolding causes rubbing at the interfaces which improves energy dissipation via friction, adding to the bulk energy already dissipated through viscoelasticity.

\section{Results}\label{sec2}

\subsection{The Unit Cell}\label{subsec2}
\begin{figure}[t]%
\centering
\includegraphics[width=\textwidth]{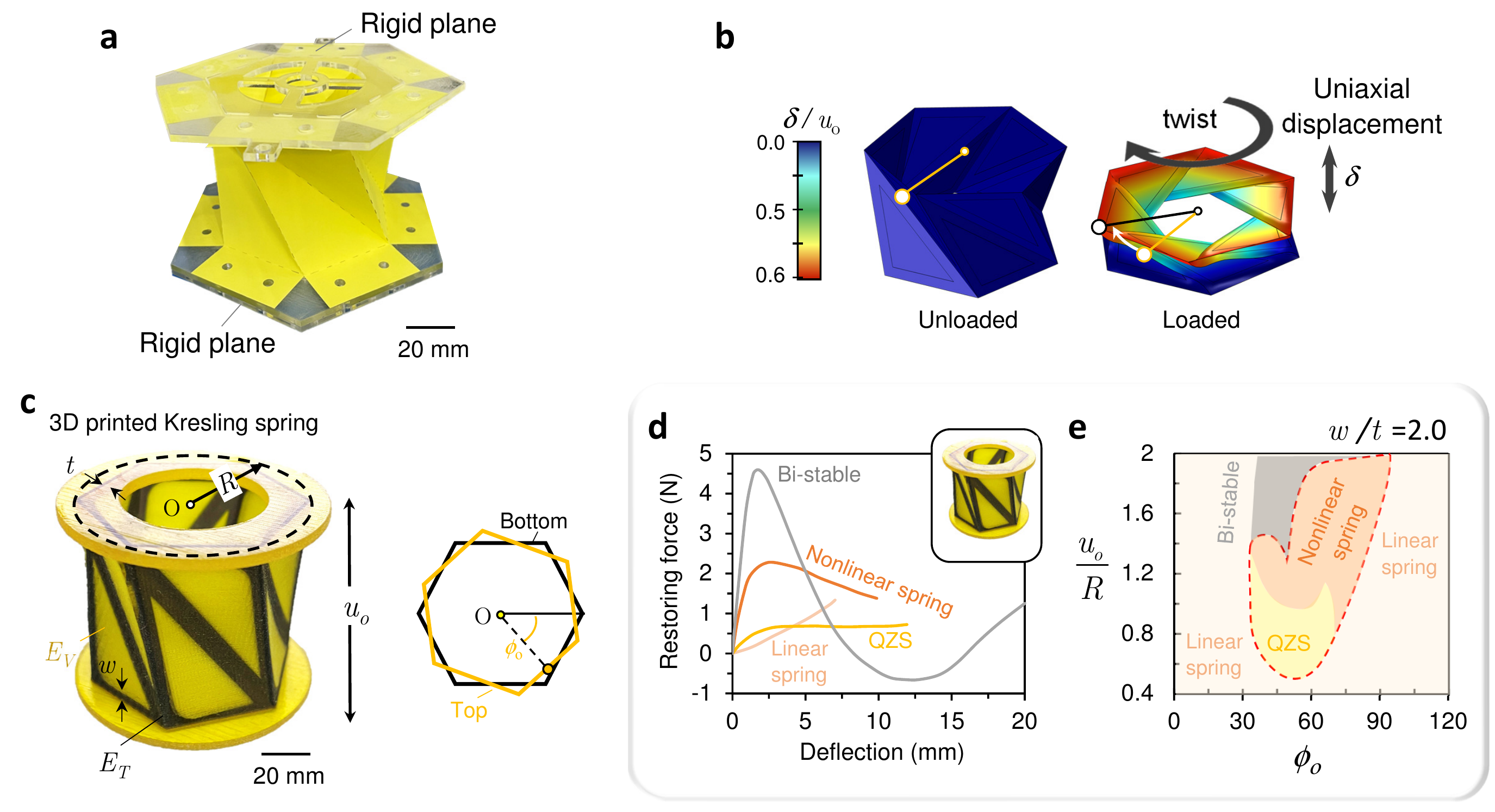}
\caption{\textbf{The Unit Cell}. \textbf{a} Paper-based Kresling Origami unit cell made by creasing and cyclical folding. \textbf{b} Deformation of the unit cell under uniaxial compression, which is characterized by simultaneous uniaxial rotation (twist) and axial deflection. \textbf{c} A 3D printed Kresling based unit cell. \textbf{d} Four distinct restoring force behaviors of the unit cell: linear, quasi-zero-stiffness (QZS), nonlinear, and bi-stable. \textbf{e} Map demarcating regions of qualitatively different  restoring force response of the unit cell as a function of the geometric parameters $\left(\frac{u_o}{R}, \phi_o \right)$. }
\label{fig2}
\end{figure}

We propose a cellular cushion material made from individual viscoelastic springs inspired by the Kresling origami pattern \cite{kresling_fifth_2020,hunt_twist_2005,noauthor_kresling-pattern_2017}. This viscoelastic spring is an adaptation of a paper-based structure constructed by folding paper following a particular procedure that involves segmenting a flat sheet of paper into triangles and folding it along the edges to form valley and mountain folds. The result is a cylindrical bellow-type structure consisting of similar triangles arranged in cyclic symmetry and connected together (Fig. \ref{fig2}a). As can be seen in Fig. \ref{fig2}b, when an external axial load or a torque is applied to the structure, it stretches or compresses depending on the direction of the applied load. In the process, the two parallel polygon planes, while remaining rigid, move and rotate relative to each other along and about a common centroidal axis similar to the motion of a threaded screw into a nut. This causes the triangular panels to deform and store the applied work in the form of strain energy at the folds. The stored energy is released upon removal of the external loads. This forces the spring to go back to its initial configuration, therewith providing a restoring element that forms the basis for the design of many interesting engineering structures.

Based on the paper spring replica, we aimed here at designing and producing a viscoelastic functional and notably durable replica of the paper-based spring that is constructed using 3D printing technologies \cite{khazaaleh_combining_2022} as depicted in Fig. \ref{fig2}c. It involves redesigning each fundamental triangle to permit folding and stretching at the panel interfaces while simultaneously providing sufficient rigidity such that it creates the desired functionality without collapsing under loading (refer to the \hyperref[FabSec]{Methods} section for details on fabrication). This is achieved by using two different materials. A compliant material is used to print the outer frames of the triangular panels, which permits stretching, bending, and extensive folding during deformation. A hard material is used to print the inner rigid cores that retain the Kresling topology during deformation.

The restoring force behavior and size of each unit cell are governed by 6 geometric and 4 material design parameters. The geometric design variables are the radius, $R$, circumscribing the top and lower polygons, the pre-deformation height, $u_o$, the pre-deformation angle, $\phi_o$, the panel thickness $t$, the flexible frame width $w$, and the number of polygon sides, $N$. The material design variables are the Young’s moduli, ($E_V, E_T$) and Poisson’s ratios ($\nu_V, \nu_T$) of the panel and creases, respectively. As shown in Fig. \ref{fig2}d,e, depending on the values of primarily $u_o/R$ and $\phi_o$, the viscoelastic spring can exhibit different restoring force behaviors, which includes a linear spring behavior, a nonlinear spring behavior, a quasi-zero stiffness (QZS) behavior, and a bi-stable behavior \cite{dalaq_experimentally-validated_2022}. The behavior is determined by the ratio $u_o/R$ and $\phi_o$ irrespective of specific values of $u_o$ and $R$. Throughout this study, the behavior of the unit cell is changed by varying $u_o/R$ and $\phi_o$ while the rest of the parameters listed above are kept constant (see Supplementary Table 1). A sufficient contrast, in terms of the Young's modulus between the constituting materials, is essential to impart enough flexibility at the creases. In addition, the material forming the creases must be stretchable (e.g., rubbery polymers). Our choice of the 3D printing materials, TangoBlackPlus (soft) and Vero (stiff), fulfills the above and provides a contrast of $E_T/E_V\approx0.0001$. The same Origami springs can be fabricated from other material combinations, with similar attributes and with comparable $E_T/E_V$ to that of Tango and Vero. 

Of particular interest to the design of cushion material is the QZS behavior because it mimics the behavior of an ideal cushion material as described earlier in Fig. \ref{fig1}c. To characterize the region/regions in the geometric design parameters space which results in a QZS response, we carried out a detailed modeling of the quasi-static behavior of the unit cell using finite element methods (FEMs) as described in the \hyperref[CompMod]{Methods} section. Figure \ref{fig2}e demarcates regions of qualitatively different restoring force behavior as function of the design parameters $\frac{u_o}{R}$ and $\phi_0$ for $N=6$. It can be clearly seen that there is a sizable region in the parameter space which leads to the desired QZS behavior.

\begin{figure}[t]%
\centering
\includegraphics[width=\textwidth]{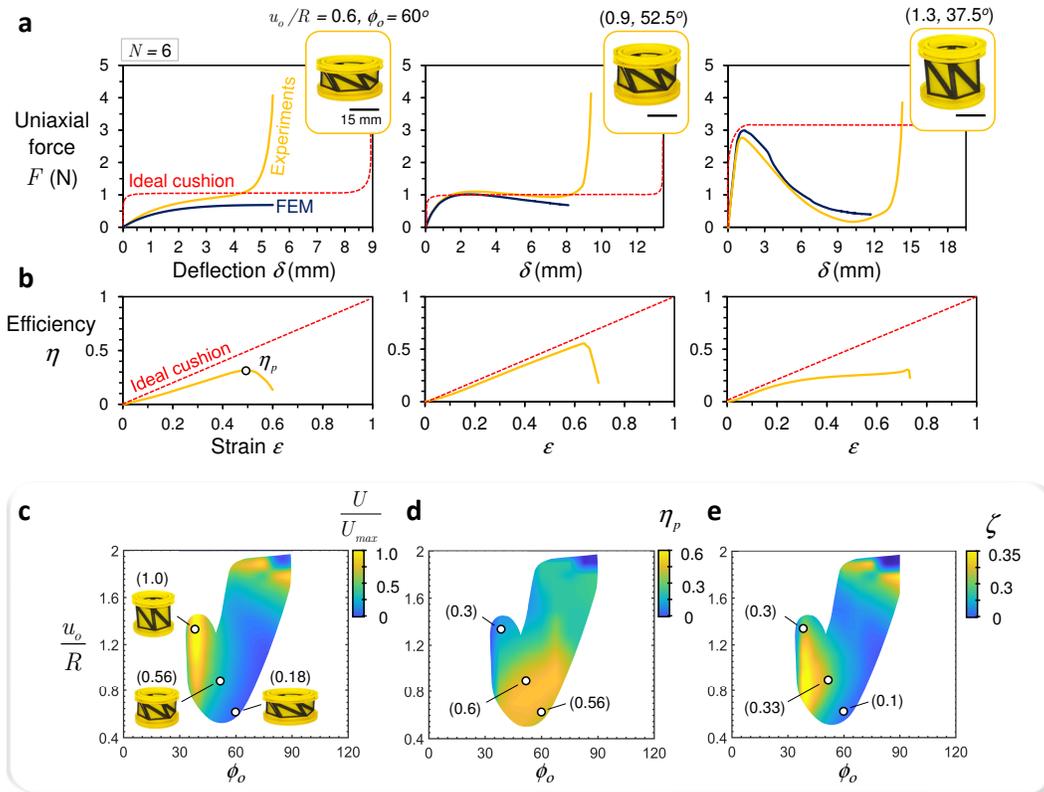}
\caption{ \textbf{Effect of geometric parameters on quasi-static response}.  \textbf{a}  $F-\delta$ and \textbf{b} $\eta-\varepsilon$ curves under quasi-static loading for three distinct designs: ($\frac{u_o}{R}=0.6, \phi_0=60^\circ$), ($0.9, 52.5^\circ$) and ($1.3, 37.5^\circ$). Computationally generated maps of  \textbf{c} normalized energy absorption $\frac{U}{U_{max}}$, \textbf{d} peak energy efficiency, $\eta_p$, and \textbf{e} combined performance indicator $\zeta=\eta_p  \frac{U}{U_{max}}$. The corresponding $U/U_{max}$, $\eta_p$ and $\zeta$ are reported between round brackets for the three designs. The reported $F-\delta$ curves are an average of three tests with an error margin of $\pm 0.02$ N. The experimental values of efficiency for the three designs are: $\eta_p=0.32\pm0.01$, $\eta_p=0.56\pm0.02$ and $\eta_p=0.30\pm0.01$, respectively.}\label{fig3}
\end{figure}

Using three different geometric data sets from the map, we generated the quasi-static $F-\delta$ curves for three different unit cells as depicted in Fig. \ref{fig3}a. Results were generated using FEM and compared to experimental data obtained using a quasi-static uniaxial compression testing machine as detailed in the \hyperref[QuasiTest]{Methods} section. The three cases are a nonlinear spring  $(u_o/R=0.6, \phi_o=60^o)$, a quasi-zero stiffness (QZS) spring $(0.9,52.5^o)$, and a strongly nonlinear softening spring $(1.3, 37.5^o)$. The first unit cell shows a nonlinear softening restoring-force behavior that is free from force fluctuations, but has a very slow rise to the steady-state value when compared to the ideal cushion response. This response is depicted by cases that have just transitioned into QZS away from the border between the QZS  (yellow region) and the linear spring response (pink region) (Fig.\ref{fig2}e). The experimental response is qualitatively similar to the computational results, with the addition of the rapid increase of force at large displacements due to densification. Note that densification due to contact between surfaces was not modeled computationally.

The second unit cell was designed using the parameters ($\frac{u_o}{R}=0.9, \phi_o=52.5^\circ$), which lies within the QZS region of Fig. \ref{fig2}e. The resulting computational and experimental $F-\delta$ curves under quasi-static conditions are similar, showing an almost QZS behavior that mimics the ideal cushion response (red dashed curve in Fig. \ref{fig3}a). The resulting response is free from overshoots and exhibits the desired flat curve (i.e., quasi-zero stiffness) after the sharp linear increase in force. The third unit cell, which is designed using ($\frac{u_o}{R}=1.3, \phi_o=37.5^\circ$), has a $F-\delta$ curve which exhibits an initial sharp increase in force followed by a strong softening behavior that ends with a spike in force due to densification. This particular design lies near the border between the nonlinear and bi-stable response (orange-grey regions) as shown in Fig. \ref{fig2}e. Increasing $u_o/R$ leads to severe softening and loss of load bearing capacity, which eventually induces bi-stability. 

\subsection{Energy Absorption Efficiency} 
One approach to measure the resemblance of the $F-\delta$ curves to that of an ideal cushion is by tracking the energy absorption efficiency, $\eta$, during deformation, which is given by \cite{li_compressive_2006}:
\begin{equation}
\eta= \frac{\int_0^{\varepsilon} \sigma (\bar{\varepsilon}) \text{ d} \bar{\varepsilon}}{\sigma_p(\varepsilon)\text{ } \varepsilon_f},
\end{equation}				       
where $\sigma$ is the engineering stress at any given engineering strain, $\varepsilon = [0, \varepsilon_f]$, $\sigma_p$ is the maximum recorded stress as $\varepsilon$ is increased, $\sigma_p=\text{max}(\sigma(\varepsilon))$, and $\varepsilon_f$ is the final strain at full compression, that is $\varepsilon_f=1$. In essence, the absorption efficiency measures the ratio of the energy absorbed up to any given deflection to that of an ideal cushion, $\sigma_p\varepsilon_f$. Therefore, $\eta$ assesses the resemblance of $F-\delta$ profile to its ideal curve (red dashed curve in Fig. \ref{fig3}a and b).  

Figure \ref{fig3}b depicts the values of $\eta$ for the three different designs as compared to the ideal cushion. Peak efficiency $\eta_p$ (indicated by a filled white circular marker) marks the onset of densification. The first and third unit cells have an efficiency that resembles that of the ideal cushion for small deformations, but that quickly deviates away from it for values of $\varepsilon > 0.2$. This indicates a gradual loss of efficiency with deformation. In contrast, the second unit cell, which is QZS, has an absorption efficiency which almost coincides with that of the ideal cushion up until the point of densification.

In addition to having a good absorption efficiency, a good energy absorbing material should have a high absorption capacity. This is characterized by the total energy absorbed, $U$, which is quantified by the area under the $F-\delta$ curve. To characterize the influence of the geometric parameters of the unit cell on these two performance metrics, we used the computational model to sweep (a full factorial sweep) across the geometric parameter space $(\frac{u_o}{R}, \phi_o)$. The sweep covered values between $\frac{u_o}{R} = [0, 2]$ and $\phi_o= [0, 120^\circ]$ with increments of 0.05 and $2^\circ$ along the $\frac{u_o}{R}$ and $\phi_o$ axes, respectively, for a total of 651 simulations. Using those simulations, we computed the total absorbed energy $U$, and the peak efficiency $\eta_p$ for each geometry. We normalized $U$ by $U_{max}$ which is the maximum value of the absorbed energy across the parameter space. Results are shown in Figs. \ref{fig3}c and d, which depict, respectively, contour maps of the normalized absorbed energy and the peak efficiency as a function of $(\frac{u_o}{R}, \phi_o)$. It is evident that the regions resulting in highest energy absorption are different from those resulting in maximum efficiency. Actually, in the region where the geometry yields the desired QZS behavior, these two performance criteria appear to be competing against each other. Thus, we devise a new performance metric which combines both criteria in the form $\zeta=\eta_p \frac{U}{U_{max}}$. High values of $\zeta$ generally indicate a better cushion. Fig. \ref{fig3}e shows a contour map of $\zeta$ as a function of the geometric parameters $(\frac{u_o}{R}, \phi_o)$. The region yielding the highest values of $\zeta$ occurs in the mid-left region between $0.5 \leq \frac{u_o}{R} \leq 1.4$ and $37^\circ \leq \phi_o \leq 50^\circ$.

\subsection{Effect of Viscoelasticity} \label{viscoelasticity}
\begin{figure}[t!]%
\centering
\includegraphics[width=\textwidth]{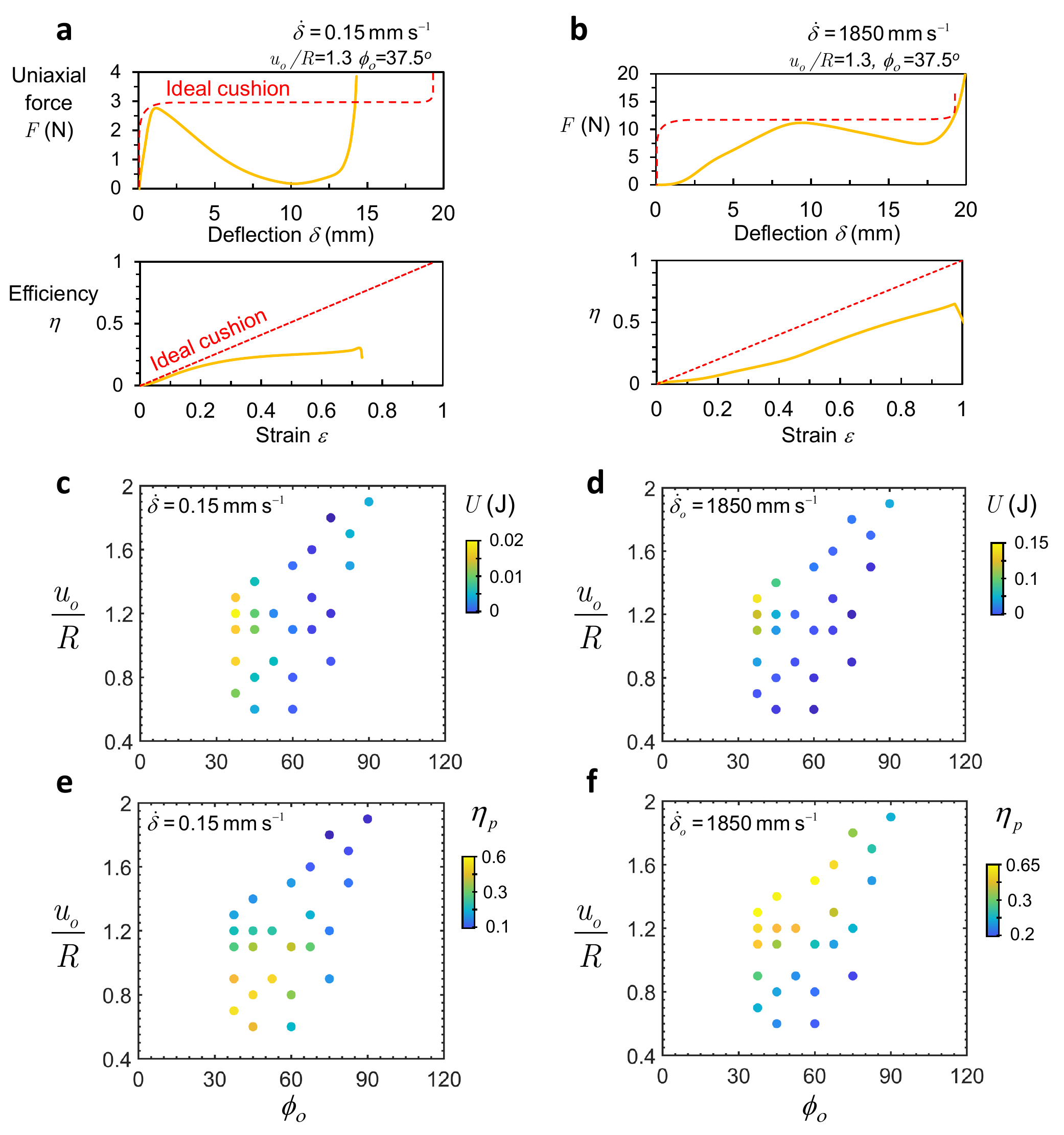}
\caption{ \textbf{Rate dependence of energy absorption and efficiency.}  \textbf{a} $F-\delta$ and $\eta-\varepsilon$ curves under quasi-static loading. \textbf{b} $F-\delta$ and $\eta-\varepsilon$ curves under impact loading. Absorbed energy $U$ and peak energy efficiency $\eta_p$ under \textbf{c,e} quasi-static and \textbf {d,f} impact loading. The absorbed energy and efficiency under quasi-static and impact loading are, respectively, ($U=0.0173\pm0.0004$ J, $\eta_p=0.30\pm0.01$) and ($U=0.146\pm0.004$ J, $\eta_p=0.65\pm0.02$).}\label{fig4}
\end{figure}

While the quasi-static $F-\delta$ curves offer a good understanding of the energy absorption capacity and efficiency of the unit cell, they provide no insight into the loading rate dependence of the $F-\delta$ curves or the amount of energy dissipated during an impact event due to viscoelasticity. First, we analyze the viscoelasticity induced rate-dependence of the $F-\delta$ curves. To this end, Fig. \ref{fig4}a and b show experimentally obtained $F-\delta$ and $\eta-\delta$ curves for the third unit cell with $(\frac{u_o}{R}=0.6, \phi_o=60^\circ)$ under quasi-static and impact loading rates of $\dot{\delta} = 0.15 \pm 0.001$ mm/s and $\dot{\delta}_o=1850 \pm 20$ mm/s, respectively. Here, we introduced the subscript ``$o$" to make the distinction that $\dot{\delta_o}$ is the initial loading rate imparted on the sample at the onset of impact. It is evident that the two curves are very different. First, the level of force experienced during high speed impact are much higher than the quasi-static case. Second, maximum force is experienced at a larger strain level during impact loading; that is, the force rises to its maximum value at a much slower rate. This has the influence of reducing the absorption efficiency at small strains. Finally, there is a reduction in the strain softening behavior observed in the quasi-static case. This causes the absorption efficiency to increase at medium and high strain levels.

To better understand which combination of geometric design parameters yields the highest absorption capacity and efficiency under impact loading, we probed the design space experimentally at 25 different locations across the geometric parameter space $(\frac{u_o}{R}, \phi_o)$. We chose points of interest to be scattered in the QZS region as encouraged by the numerical simulations shown in Fig. \ref{fig3}d. We also included points around the QZS region pertaining to the nonlinear-softening cases, as they may have the potential of exhibiting QZS behavior upon increased loading rate due to viscoelasticity. For each geometry, we experimentally measured the $F-\delta$ and $\eta-\delta$ curves, which, in turn, are used to compute $U$ and $\eta_p$ as shown in Figs. \ref{fig4}c and e for the quasi-static case, and Fig. \ref{fig4}d and f for the impact scenario. Each point on the figures corresponds to the average of 3 tests. For the quasi-static case, best performing designs in terms of energy absorption, $U$, cluster around the left-hand side of the parameter space, which agrees well with the computational simulations shown earlier in Fig. \ref{fig3}d. The best performing cluster in terms of energy absorption, $U$, remains the same even under impact loading. Thus, one can safely rely on quasi-static simulations to identify the best energy absorbing designs regardless of loading rate.

In terms of absorption efficiency, best designs under quasi-static testing cluster around the bottom left corner of the figure favoring shorter and more compliant unit cells (Fig. \ref{fig4}e). On the other hand, best performance under impact loading favors longer designs with larger stroke distance as can be clearly seen in Fig. \ref{fig4}f. Moreover, when comparing Figs. \ref{fig4}c and d, we notice that, due to the viscoelasticity of the base constituents, samples absorbed 7.5 times more energy under impact loading than those subjected to quasi-static loading. 

Next, we analyze the effect of the loading rate on the dissipated energy, which can be captured by quantifying the size of the hysteresis loops during a single cycle of loading and unloading. To this end, we performed loading tests at five different loading rates: $\dot{\delta}$ = 0.1, 1, 10, 20 and 50 mm/s and tracked the size of the resulting hysteresis loops (Supplementary Note 1, Supplementary Fig. 3a). It is evident, that the size of the hysteresis loop corresponding to the energy dissipated per cycle increases with the loading rate (Supplementary Note 1, Supplementary Fig. 3b). This hysteresis is consistent and notably predictable (see also Supplementary Movie 1). Thus, unlike energy dissipation via damage, viscoelasticity provides a predictable energy dissipating mechanism which improves with the loading rate. 

\subsection{The 3D Tessellated Lattice}
\begin{figure}[t]%
\centering
\includegraphics[width=\textwidth]{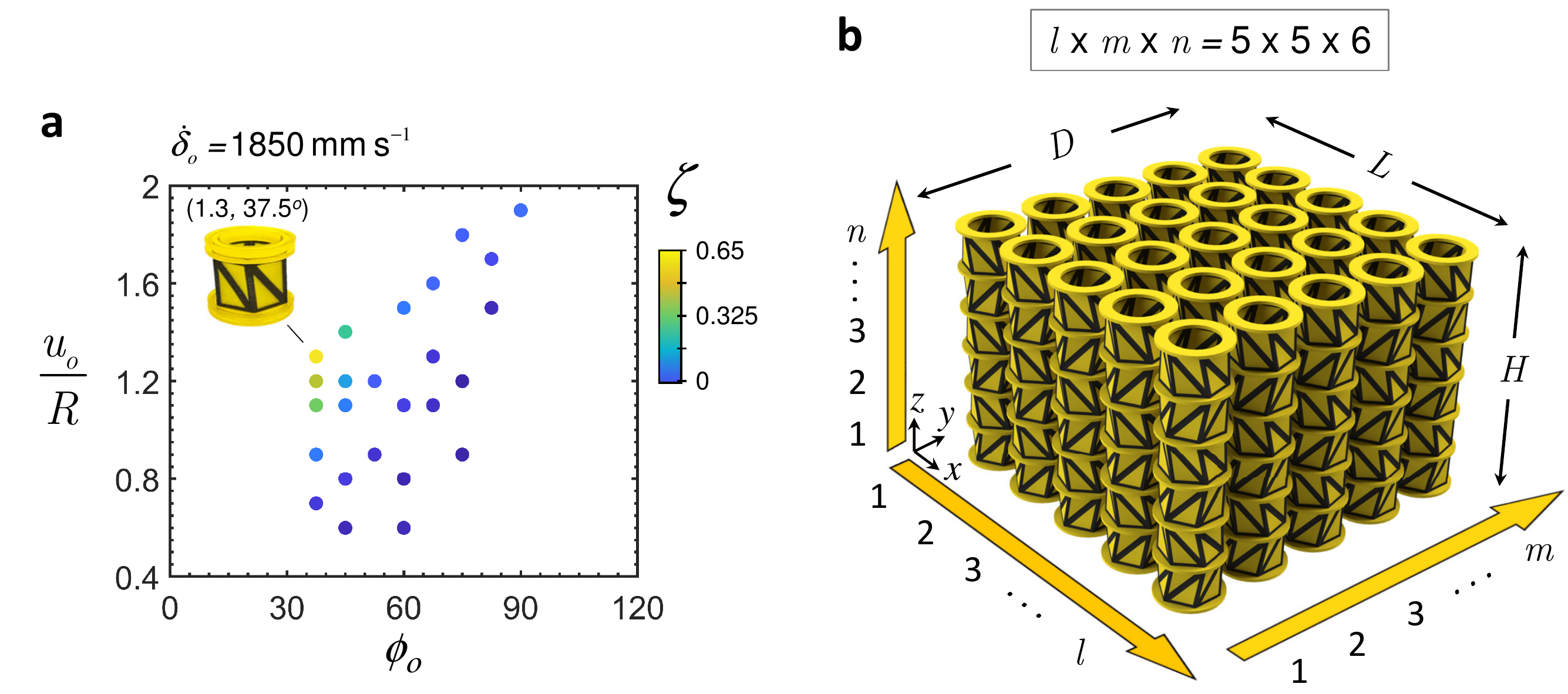}
\caption{ \textbf{Optimal unit cell design}.  \textbf{a} A map showing experimental values of the combined performance indicator $\zeta$,  based on impact testing. \textbf{b} Architectured cellular cushion made up of tessellation of $l \times m \times n$ unit cells along the $x, y$ and $z$ directions. For the optimal design $(1.3, 37.5^o)$, the performance indicator is $\zeta=0.65\pm0.06$.}\label{fig5}
\end{figure}

The combined performance indicator $\zeta$ is used to identify the optimal unit cell design for impact absorption. Experimental measurements in Fig. \ref{fig5}a show that the best performing design in terms of both energy absorption and efficiency corresponds to the geometry with ($\frac{u_o}{R}=1.3, \phi_o=37.5^\circ$). We therefore use this optimal unit cell to architecture and construct, for the first time, a 3D Origami based composite lattice which is depicted in Fig. \ref{fig5}b. This design utilizes alternating chirality to cancel out any net rotation during compression of the springs. This design which is created by tessellating the unit cell along the three Cartesian directions: $l = 5$, $m = 5$ and $n = 6$, has a total of 150 unit cells. 
 
To first understand the mechanics of such tessellated designs, we simulated their responses for series and in-plane tessellations. Tessellation of unit cells in series along the $z$-axis forms a vertical stack, or rather an Origami column of $1\times 1\times n$ unit cells, with an equivalent height of $H =nu_o$. Since equilibrium along the  $z$-axis requires that the reaction uniaxial force, $F$, be constant and independent of $n$, taller columns with larger height $H$ are more compliant and have larger stroke distance. The larger stroke distance enhances energy absorption (area under the $F-\delta$ curves), which directly scales with the number of vertical unit cells as shown in Fig. \ref{fig6}a for $n = 1$ to $5$ (see Eq. (22) in Supplementary Note 3). 

Figure \ref{fig6}b shows the field plots of the deforming Origami-column at $\frac{\delta}{u_o} =0, 0.5$ and $0.6$ for both $n = 3$ and $4$. The junctions between the unit cells are indicated by $\Omega_i$, where the subscript $i$ denotes the $i^{th}$ junction location. For example, a column of $n=4$ has 5 junctions as indicated in Fig. \ref{fig6}b. Those plots are obtained for uniaxially compressed columns subjected to uniaxial displacement at top surface and that is fixed at the bottom for all cases (\hyperref[CompMod]{Methods} section). Since unit cells are stacked with alternating opposite chirality, no rotation occurs at the ends of each pair. Instead, the junction in between the two pairs (even values of  $i$) absorbs the entire rotation. Rotations are therefore induced at every other junction. For instance, for $n=3$, the compression at the top, junction $\Omega_4$, induces rotation at $\Omega_4$ and $\Omega_2$, while $\Omega_3$ and $\Omega_1$ are free of any rotation. 

Tessellation within the $x-y$ plane forms a planar lattice of $l\times m \times 2$ unit cells. Top and bottom unit cells are connected by rigid planes to ensure direct transfer of forces across the model. The simulated $F-\delta$ curves for $l\times m= 5 \times 1$, $2 \times 2$, $3 \times 1$, $2\times 1$ and $1\times 1$ are shown in Fig. \ref{fig6}c for $n=2$. It is evident that the overall response becomes stiffer as $l\times m$ increases, requiring a larger uniaxial force $F$ to achieve a unit deflection. The total energy absorbed $U$ scales directly with the total number of in-plane unit cells, i.e. $U \propto l \times m$ (see Supplementary Note 3). 

\subsection[]{The $F-\delta$ curve of the 3D Tessellated Lattice}
\begin{figure}[t]%
\centering
\includegraphics[width=\textwidth]{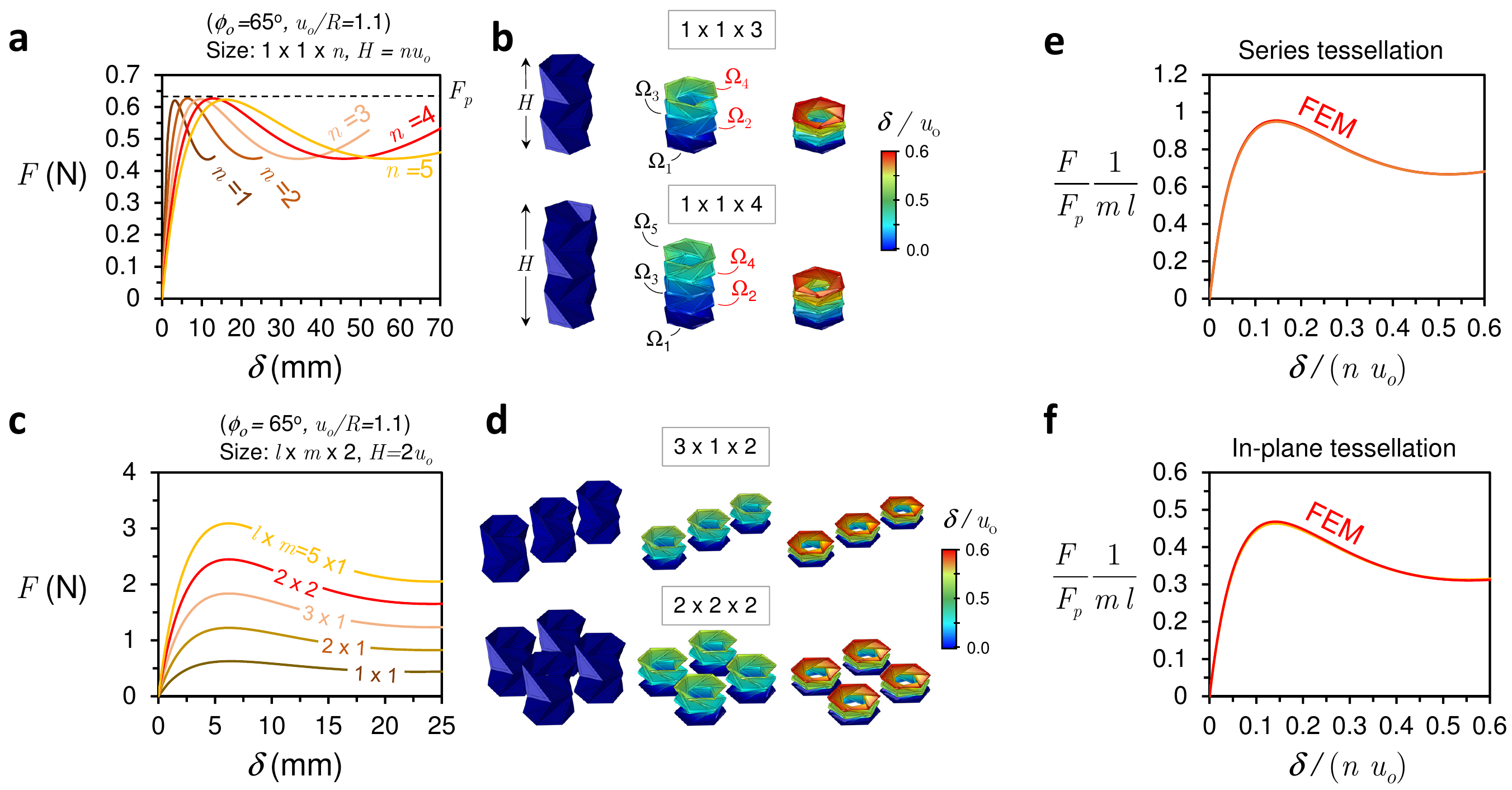}
\caption{ \textbf{Mechanics of tessellation}. \textbf{a} $F-\delta$ curves for a series tessellation as function of the number of tessellated unit cells $n$. \textbf{b} Deformation of a vertically tessellated column for odd ($n=3$) and even ($n=4$) number of cells. \textbf{c} $F-\delta$ curves for in-plane tessellation using $l\times m \times 2$. \textbf{e,f} Universal $F-\delta$ curves for series and in-plane tessellation of unit cells.}\label{fig6}
\end{figure}

While FEM can be used to generate the $F-\delta$ curves of the tessellated lattice as shown in Fig. \ref{fig6}a and c, more insights can be gained form an analytical description of those curves. Similar to hyperelastic foams, we model each unit cell as a nonlinear hyperelastic material  (Supplementary Fig. 6a). The $F-\delta$ curve of a single unit cell can then be modeled by adopting Storaker equation for hyperelastic materials \cite{storakers_material_1986} (Supplementary Note 3):

\begin{equation}
\frac{F_c}{k_c u_o}=\frac{1}{a_c+b_c}\left\{\left(1-\frac{\delta}{u_o}\right)^{a_c}-\left(1-\frac{\delta}{u_o}\right)^{-b_c}\right\},
\label{Fc}
\end{equation}
which can be extended to account for tessellation in 3D using
\begin{equation}
\frac{F}{k_c H}=\frac{l m}{n}\left(\frac{1}{a_c+b_c}\right)\left\{\left(1-\frac{\delta}{H}\right)^{a_c}-\left(1-\frac{\delta}{H}\right)^{-b_c}\right\},
\label{F}
\end{equation}
where $F_c$ is the force experienced by the unit cell; $k_c$ is a constant representing the stiffness of the material at $\delta=0$; $a_c$ and $b_c$ are empirical constants that can be calibrated based on either experiments or high-fidelity simulations of a single unit cell. Those constants reflect the particularities of the geometric and material properties of the cell. Equations (\ref{Fc}) and (\ref{F}) can be used to fully capture the restoring force behavior of any single unit cell and the overall response of any lattice of known configuration, $l\times m \times n$. Results showing the agreement between this analytical model and experiments can be found in Supplementary Figs. 6b and 6c. 

The analytical model presented in Eq. (\ref{F}) offers a key additional insight that cannot be easily inferred by relying on the FEM simulations. In particular, by introducing the dimensionless deflection $\delta/(nu_o)$ and the dimensionless force $F/(m l F_p)$, where $F_p$ is the peak force in any given $F-\delta$ curve, all the curves in Figs. \ref{fig6}a and c can be collapsed into two universal $F-\delta$ curves for series and in-plane parallel tessellations. This facilitates the prediction of the $F-\delta$ curve for any arbitrary tessellated lattice.

Combining the series and parallel configurations offers a rich platform for tuning the mechanical properties and customizing the overall $F-\delta$ curve for the desired application. The series configuration (varying $n$ only) can be used to tune the stiffness, stroke distance, and, therewith  energy absorption $U$, while keeping the peak forces fixed (constant strength). Whereas, in-plane tessellation (varying $l$ and $m$) can be used to tune the maximum force (strength), stiffness, energy absorption, and the footprint area of the cushion material. Thus, in a nutshell, we proposed here an Origami based design framework that allows tailoring of the geometric parameters of the Kresling topology ($u_o/R$, $\phi_o$) to engineer the desired $F-\delta$  response, then use it to form a 3D lattice by setting $l$, $m$ and $n$ for the desired levels of stiffness, strength, stroke distance and energy absorption.

\subsection{A Functional Cushion} 
\begin{figure}[t]%
\centering
\includegraphics[width=0.75\textwidth]{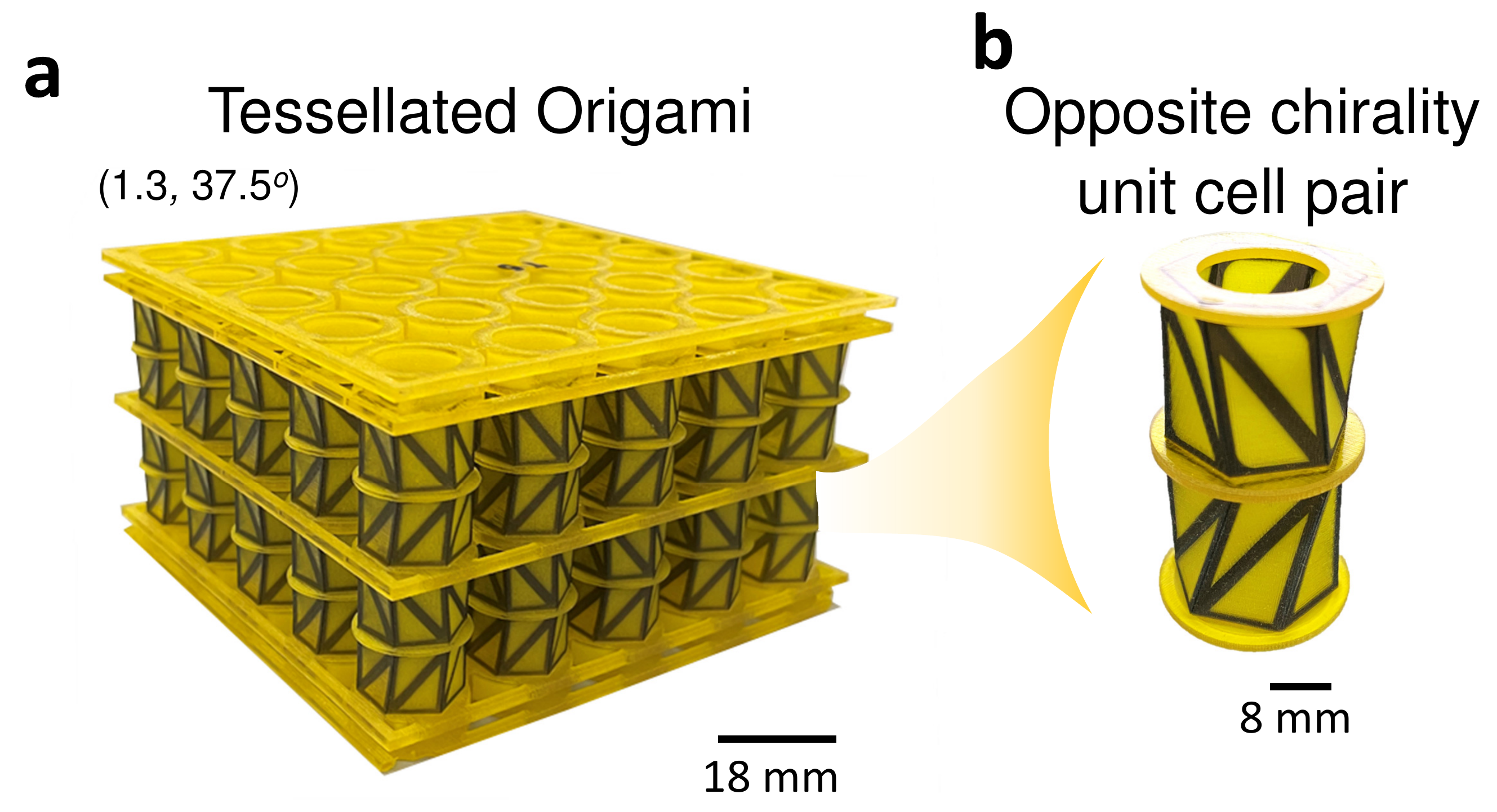}
\caption{ \textbf{Functional 3D printed Origami-based cushion}. \textbf{a} 3D printed tessellated ($l \times m \times n = 5\times 5 \times 4$) Origami-based material using the geometric parameters of ($u_o/R=1.3$, $\phi_o=37.5^\circ$). \textbf{b} A pair of opposite chirality unit cells.}\label{fig7}
\end{figure}

There have been several methods for 3D fabrication of Origami structures ranging from the use of flexible linkages, scoring sheets using laser, and paper folding \cite{ishida_design_2017,yasuda_origami-based_2017, deleo_origami-based_2020, butler_highly_2017}. Here, we employ state-of-the-art additive manufacturing utilizing the polyjet technique. By carefully tuning the UV light exposure time, printing speed, unit cell spacing and minimum feature size, we are able to 3D print a high-fidelity tessellated prototype in ``one go''. In other words, we bypass the assembly step that is often employed in other forms of tessellated metamaterials. Figure \ref{fig7}a and b show, respectively, the 3D printed cushion material using the optimal unit cell design ($\frac{u_o}{R}=1.3, \phi_o=37.5^\circ$) and the opposite-chirality unit cell forming the lattice.    
 
\begin{figure}[t]%
\centering
\includegraphics[width=\textwidth]{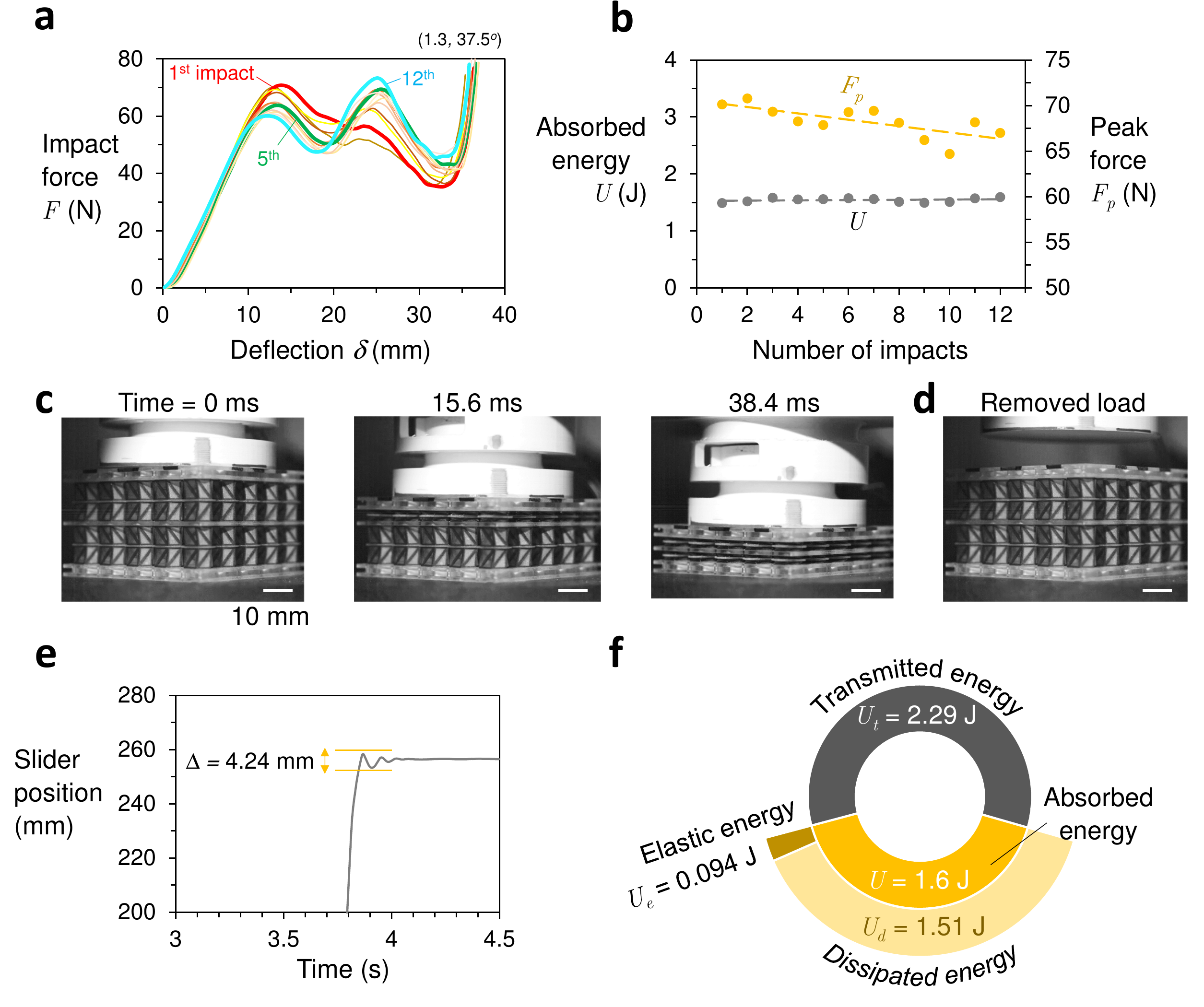}
\caption{ \textbf{Impact test of the cushion}. \textbf{a} Force-deflection ($F-\delta$) generated by impact testing of a $5 \times 5 \times 4$, 3D printed cushion with geometric parameters of ($u_o/R=1.3$, $\phi_o=37.5^\circ$), subjected to 12 cycles of impact. \textbf{b} Changes in energy absorbed $U$ and peak force $F_p$ with respect to number of impacts. \textbf{c} Snapshots of high speed imaging at three time instants: $t=0$ ms, $15.6$ ms and $38.4$ ms. \textbf{d} A snapshot of high speed camera after load removal showing full shape and height recovery. \textbf{e} Slider position vs time, showing rebound distance $\Delta$ after impact. \textbf{f} Impact energy breakdown showing transmitted energy $U_t$, elastic energy $U_e$, dissipated energy $U_d$ and absorbed energy $U$.} \label{fig8}
\end{figure}

The 3D printed prototype serves as a functional, viscoelastic, Origami-inspired cushion for impact absorption. To test its performance, we carry out a low-speed impact test on this material ($\dot{\delta} = 1850 \pm 20$ mm/s and impactor mass, $m = 2.26 \pm 0.001$ kg). In particular, we subject the material to 12 successive impacts with a window of 7 min break.  Figure \ref{fig8}a shows the corresponding $F-\delta$ curves for the multiple impacts. The first impact exhibits the highest peak force, which appears to drop slightly with each subsequent impact up to the $5^{th}$ impact, beyond which the $F-\delta$ curve is unaffected. This can be seen in Figure \ref{fig8}b, which studies variation of the peak force, $F_p$, and absorbed energy, $U$, as a function of the number of impacts. Successive impacts only cause around 5\% reduction in the peak force (strength) with no measurable influence on the overall energy absorption capacity of the cushion. 

The damage resistance of the cushion can be attributed to the unique topology of the Origami cell which channels the incident energy into graceful folding, crease stretching, and panel rubbing through rotation (see Supplementary Movie 2). The $F-\delta$ curves of Fig. \ref{fig8}a exhibit a plateau region characterized by a wavy profile which is attributed to the successive engagement of the layers in the structure. High speed imaging performed at 10,000 frames/s reveals this successive engagement of the layers as shown in Fig. \ref{fig8}c. The impactor head reaches the material at $t= 0$ ms. Within the next 15.6 ms, the head presses through the 22 mm height of the first layer, which is reflected as a linear rise in the reaction force until the peak force is reached followed by slight softening. After which, the impactor presses through the second layer exhibiting the same rise until a second peak force is realized followed by a softening behavior (see Supplementary Movie 2). It takes the impact head a much longer time of 23 ms to flatten the second layer because it has already decelerated and lost a large part of its kinetic energy. The lower impact speed reduces the viscoelastic effect, which is reflected by the higher softening response in the second peak of Fig. \ref{fig8}a. By virtue of the Kresling topology, the structure recovers 100\% of its shape and height despite multiple impacts (Fig. \ref{fig8}d). In contrast to other cellular materials, the wavy profile is not resulting from material damage, but is an inherent characteristic of the Kresling topology that is fully customizable (deterministic).

The 2.26 kg impactor approaches the Origami structure at a speed of 1.85 m/s and imparts a total energy of $3.87 \pm 0.03$ J on the cushion. Using a laser sensor, we track the impactor position during testing as depicted in Fig. \ref{fig8}e. This permits carrying out a full energy analysis to provide an energy breakdown of the impact event as shown in Fig. \ref{fig8}f. Almost 41\% ($U = 1.6 \pm 0.02$ J, area under the $F-\delta$) of the 3.9 J incident energy gets absorbed by the structure, of which 94\% ($U_d = 1.51 \pm 0.02$ J) is dissipated through viscoelasticity and friction between the panels. About 5\% ($U_e = 0.094 \pm 0.001$ J) is stored as elastic energy within the structure, and hence, is recovered as a rebound of $\Delta = 4.24 \pm 0.01$ mm. The remaining $U_t = 2.29 \pm 0.05$ J of the total energy is transmitted to the surrounding via acoustic and elastic waves. 


\subsection{Apparent and Effective Properties of the Cushion} 
\begin{figure}[t]%
\centering
\includegraphics[width=\textwidth]{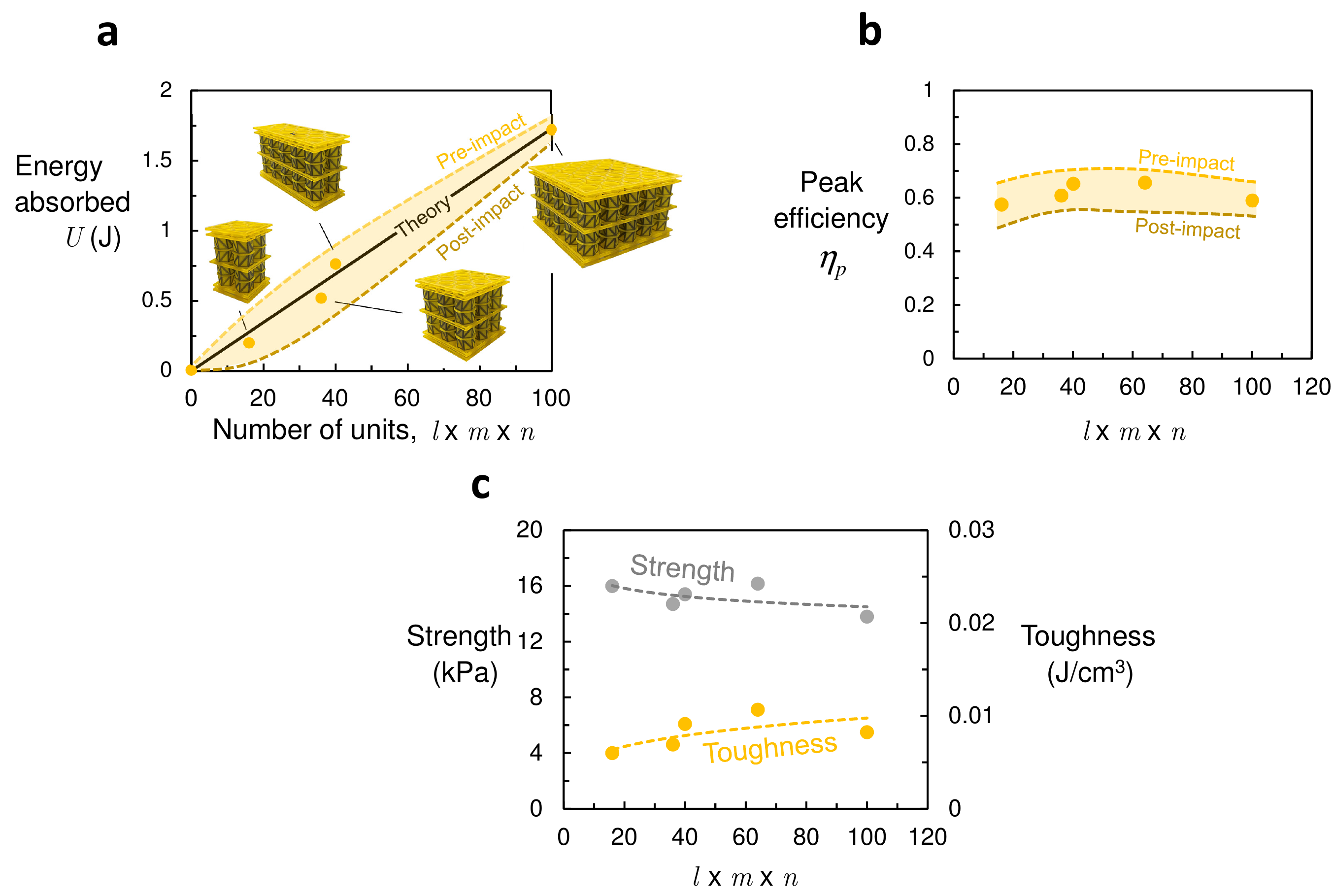}
\caption{ \textbf{Mechanics of the cushion}. Variation of \textbf{a} energy absorbed $U$, \textbf{b} peak efficiency $\eta_p$, and \textbf{c} effective strength and toughness as function of the number of unit cells $l\times m \times n$. Filled circular markers represent average experimental data.} \label{fig9}
\end{figure}

The Origami-based lattice can be either treated as a structure or as a material in its own right. Thus, it is critical to understand whether the energy absorbed via a single unit cell scales linearly with the number of unit cells. Figure \ref{fig9}a shows the raw energy absorption for $l\times m\times n = 1\times 1\times 1$, $2\times 2\times 4$, $3\times 3\times 4$, $5\times 2\times 4$ and $5\times 5\times 4$. The shaded envelope represents the experimental variation across the tested samples. Each point presented on the figure is an average value of 15 impact cycles, where the upper and lower limits of the envelope denote the pre-impact (fresh samples) and the minimal post-impact results, respectively. Knowing that the energy absorbed by a single unit cell is $U_c= 0.0173 \pm 0.0004$ J, we can predict the overall response of the lattice by using (Supplementary Note 3): 
\begin{equation}
U= (l m n) U_c
\label{U}
\end{equation}					
Equation (\ref{U}) is plotted together with the experimental data in Fig. \ref{fig9}a as a black solid line showing very good correspondence. 

We also calculate the peak efficiency, $\eta_p$, for the different samples as a function of $l \times m \times n$ as depicted in Fig. \ref{fig9}b. The absorption efficiency remains nearly constant, ranging between 0.55 - 0.7. Thus, it is plausible to conclude that $\eta_p$ is independent of the number of unit cells, and that it is a property of the geometry of each unit cell.

In view of this design framework as a cushion material, we calculate the apparent strength and toughness of the cushion (see Supplementary Note 1). Both the strength and toughness are raw measurements that are normalized by their effective area and volume, respectively. Therefore, they are independent of size and will mainly be dependent on experimental conditions such as impact speed, base material constituents, and ambient conditions. It can be seen in Fig. \ref{fig9}c that the proposed cushion achieves periodicity, and therewith its effective material properties, at a relatively low number of unit cells. Actually, only ~30 unit cells are needed for both strength and toughness to saturate with the increased number of unit cells. 

\section{Discussion}

Throughout this paper, it has been  demonstrated that origami-inspired unit cells based on the Kresling pattern can be used to architect efficient cushion materials for impact energy absorption and dissipation. Those materials, which absorb energy through graceful folding at the interfaces of each unit cell in the absence of permanent damage induced by irreversible buckling, can fully recover once the load is removed (see Supplementary Movie 3). Indeed, results illustrate that even after as many as twelve impacts that fully compress the cushion, the strength was only reduced by 5\% with no measurable influence on the overall energy absorption capacity of the cushion.

Because of the sensitive dependence of the $F-\delta$ curve on the geometric properties of the unit cell, each cell can be designed to retain a desired level of energy absorption efficiency, while tuning for specific combinations of peak force, total energy absorbed, or effective strength and toughness. This permits the design of optimal cushion materials for the application at hand (impact level and speed). 

Energy dissipation, which is mainly due to viscoelasticity and frictional rubbing between the panels, is rate and load dependent. A cushion material consisting of two layers of unit cell pairs (4 cells across the height) was able to absorb 41\% of the energy imparted by a load moving with a linear momentum of 4.2 kg m/s at a speed of 1.85 m/s, and dissipate 94\% of absorbed energy. Both percentages could possibly be increased by increasing the number of layers.

\begin{figure}[t]%
\centering
\includegraphics[width=1.0\textwidth]{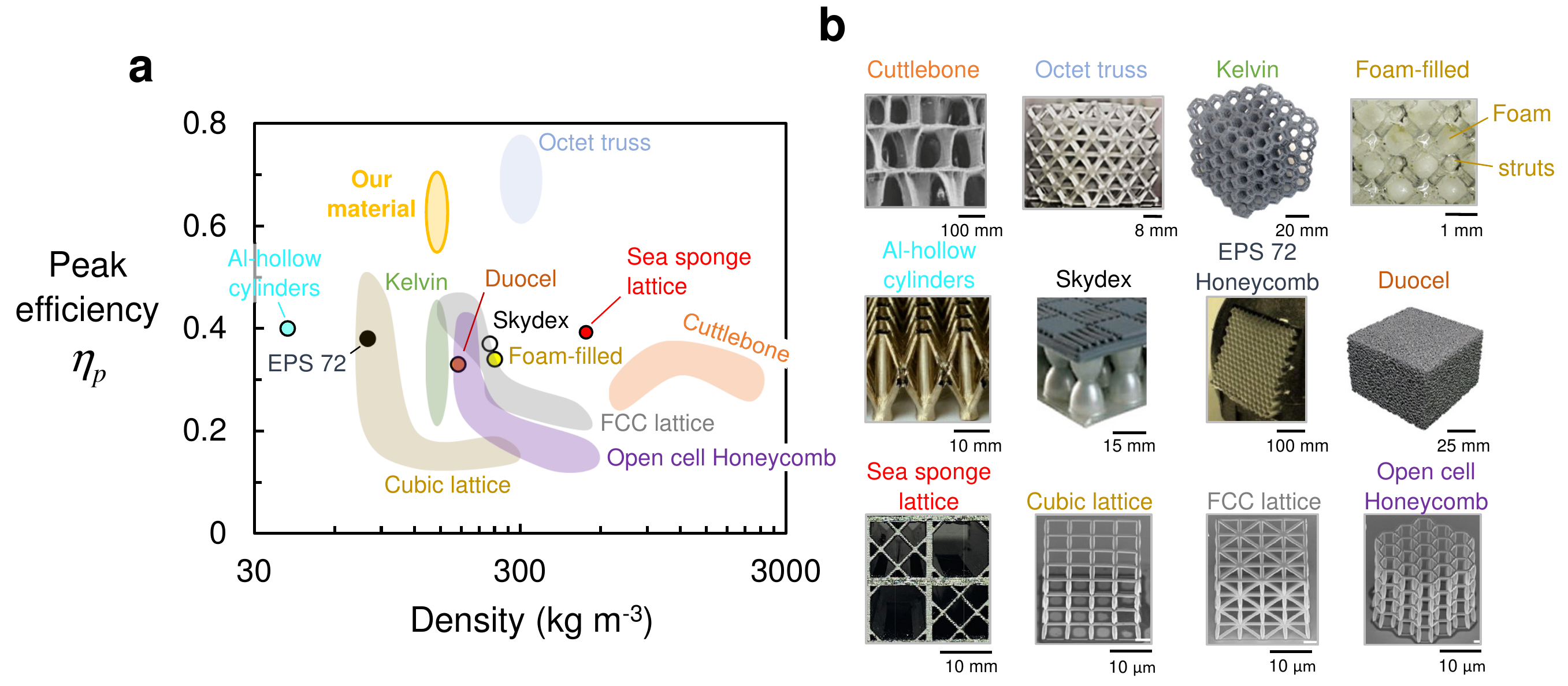}
\caption{ \textbf{Comparison chart}. (a) Efficiency-density  ($\eta_p - \rho$) chart for a number of available cushion (commercial or literature). (b) Actual images of reported lattice designs: Aluminum hollow cylinders} \cite{liu_dynamic_2014}, Aluminum Duocel \cite{mohsenizadeh_additively-manufactured_2018} (image adapted from \cite{duocel_aluminum_2023}), Open cell Honeycomb-, FCC-, cubic-\cite{mieszala_micromechanics_2017}, Kelvin-lattice \cite{habib_fabrication_2018}, EPS 72 foam (honeycomb lattice) \cite{caserta_shock_2011}, Octet-truss \cite{song_octet-truss_2019}, Skydex (twin-spherical lattice) \cite{mohsenizadeh_additively-manufactured_2018, de_sousa_assessing_2012} (Skydex padding image adapted from \cite{zhu_parameterized_2014}), Foam-filled (COF240) \cite{ramirez_viscoelastic_2018}, Sea sponge lattice \cite{sharma_bio-inspired_2022} and Cuttlebone-lattice \cite{yang_advanced_2020} (image adapted from \cite{cadman_cuttlebone_2012}). The Kresling based Origami material is denoted by ``Our material'' in orange. Unless otherwise stated, images are adapted from their respective reference(s), from which numerical data is reported. \label{fig10}
\end{figure}

In order to compare the performance of this cushion material to other existing materials or lattices structures, we generate an absorption efficiency versus density  ($\eta_p - \rho$) chart as shown in Fig.  \ref{fig10}. Some of those reported cases are regarded as structures as they do not achieve separation of scale, where the magnitudes of the forces or apparent Young's modulus are still scale dependent. Recall that the peak efficiency, $\eta_p$, assesses the closeness of $F-\delta$ curve to its ideal response (red curves in Fig. \ref{fig3}). As such, it primarily reflects the effect of geometry and viscoelasticity (loading rate dependence, if any) on the overall response irrespective of scale or specific Young's modulus of the lattice's constituents. The ideal cushion materials must be able to provide the highest absorption efficiency at the lowest possible density. The Origami base unit cell has a panel thickness of 0.4 mm and the material has a small apparent density $\rho = 138$ kg/m$^3$ leading to a light and highly porous cellular material with a relative density of  $\rho/ \rho_s= 0.12$ ($\rho_s$ is the density of solid material). 

In terms of absorption efficiency, the vast majority of extensively researched cellular topologies used for the design of architectured cellular cushions; e.g. cubic-, kelvin-, honeycomb- and FCC-lattices, have an energy absorption efficiency in the range $0.15<\eta_p<0.5$. Industrial counterparts, such as Ducel, Skydex, EPS 72 foams, have a peak efficiency which hovers around  $\eta_p =0.4$. Notably, the Origami-inspired cushion demonstrated a high energy absorbing efficiency of around $\eta_p =0.7$ at very low density. In particular, experimentally collected data yields efficiency numbers that occupy the top left corner of the chart,  $0.6<\eta_p<0.7$. In addition, because of the viscoelasticity-induced dependence of the absorption efficiency on the  rate of impact, efficiency figures are expected to increase even further at higher impact speeds. Only architectured materials based on the  Octet truss lattice have a comparable efficiency  under quasi-static conditions. However, this material has a much higher density and suffers substantial damage under impacts with low recoverability of its initial height, shape, and structural load bearing capacity. In contrast, the Origami-inspired cushion is damage resistant, recovers its full shape, and has a reliable mechanical response even after a large number of impact events. Finally, a solution to small impactors or incident bodies interacting with the through-holes of the cushion can be achieved by incorporating a top rigid plate containing arrays of tiny perforations to allow air to escape during impact while preventing the interactions with the holes (Supplementary Fig. 7). The plate can also be made flexible to selectively localize the impact and control the participation of the unit cells during impact.

\section{Materials and Methods}\label{sec11}

\subsection{Fabrication}\label{FabSec}
The Kresling unit cell and its full 3D tessellated version are fabricated using Stratasys J750 3D printer (Stratasys \cite{stratasys_feel_2022}, Commerce Way, USA) which utilizes the polyjet technique. The structure is made out from a compliant (rubbery) and hard materials; namely, TangoBlackPlus \cite{stratasys_tango_2018,object_fullcure_2009} and Vero \cite{stratasys_vero_2018, object_fullcure_2009} (Fig \ref{fig2}c). The compliant material is used to create the creases, particularly outer frame of the unit cell to permit stretching, bending, and extensive folding during deformation, while the hard material is used to create the inner rigid core, which retains the Kresling topology during deformation. Each 3D printed sample is submerged in a chemical solution consisting of 2\% caustic soda (sodium hydroxide) and 1\% sodium metasilicate (Na$_2$SiO$_3$) for a period ranging from 18 to 72 hours, depending on the size of the sample, to ensure gentle and complete removal of the SUP706 support material \cite{stratasys_find_2022}. Each sample  is then cured under UV-light for 300 minutes. In practice, the size of the tessellated Origami is limited by the size of the build plate of the 3D printer, which is $490 \times 290 \times 200$ mm. For mass production, full utilization of the build plate allows the printing of 12 cushion samples (of $l \times m \times n = 5\times 5 \times 4$, Fig. \ref{fig7}a), or a full-fledged tessellation over the entire plate of $60\times 60 \times 4$, in 42 hours. Those 12 samples of $5\times 5 \times 4$ consume a total of 295 grams of TangoBlackPlus, 1403 grams of Vero and 8600 grams of support materials (SUP706). We can thus fabricate, clean and cure samples in $\sim65$ hours (or an average of $\sim5.42$ hours per a single $5\times 5 \times 4$ cushion).

\subsection{Uniaxial Testing}\label{QuasiTest} The uniaxial quasi-static tests are performed using an Instron 5960 series universal testing machine (Instron \cite{illinois_tool_works_inc_out_2020}, Norwood, US). The sample is placed into a sample holder designed in-house as shown in Supplementary Fig. 1a. The bottom part of the sample holder is connected to a frictionless bearing which permits a smooth rotation of the unit cell upon compression. The entire apparatus rests on a rigid platform. A constant rate displacement of 0.15 mm/s is applied at the upper surface of the holder up to 80\% of the initial height. The resulting axial reaction force, $F$, exerted by the sample on the clamp is recorded using a load cell.  The measured deflection and reaction force are used to generate the $F-\delta$ curves (Supplementary Fig. 1b). The resulting curves are then translated into their corresponding engineering stress-strain and efficiency-strain curves \cite{li_compressive_2006} (see Supplementary Note 1 for further details on the uniaxial testing process).

\subsection{Computational Model}\label{CompMod} A general shell-based finite element model is built using Abaqus (Dassault Systèmes \cite{simulia_abaqus_2022}, US) to simulate the quasi-static behavior of the Origami-based cellular materials. The base constituents are considered to be linear-elastic materials. The model is meshed using 3D planar shell elements that account for 3D translation, rotation, in-plane and out-of-plane deformations \cite{michael_smith_abaqusstandard_2019}. The model takes the geometric parameters $(\frac{u_o}{R}, \phi_o, w/t)$, the material elastic properties ($E_T, \nu_T, E_V, \nu_V)$, and tessellation ($l \times m \times n$) as inputs and uses them to construct the 3D shell-model using the Kresling topology. Considering an Origami column ($1 \times 1 \times 4$) as an example (Supplementary Fig. 4), the bottom surface $\Omega_1$ is fixed while the top surface $\Omega_{n+1}= \Omega_5$ is subjected to uniaxial compression $u_z =-\delta$. To ensure a well-posed boundary value problem with unique solutions \cite{kim_introduction_2008} and that is free of numerical artifacts, all junctions, including top and bottom surfaces ($\Omega_1 - \Omega_{n+1}$), are constrained such that their rotation around the $z$-axis is restricted for odd junctions. In other words, the rotation at every other two junction(s) is constrained. Otherwise, any subtle variation in the relative displacement/rotation among junctions in the course of the numerical simulation results in junctions overtaking or falling behind during deformation, which is solely due to numerical artifacts. We also restrict the out-of-plane rotation (overturn) of all junctions. All remaining surfaces and edges are assumed to be traction free. The above constraints are imposed through the boundary conditions listed in Supplementary Note 2 and in Supplementary Fig. 4.

\subsection{Impact Testing} 
The impact testing rig consists of a free-falling impactor head guided via two frictionless bearings along two steel poles. The unit cell sample is positioned under the impactor head with its bottom surface clamped to a frictionless bearing to allow rotation under load (Supplementary Fig. 5a). For the tessellated cushion, a bearing is not required since its end surfaces are free of rotation. The mass of the impactor head can be adjusted by adding different weights via a bolt-nut mechanism. The mass of the slider and the attached masses is 2.26 kg for impact tests involving the tessellated cushion material. Whereas, the total mass is set to 1.26 kg for the single unit cell impact tests. A laser sensor (Micro-Epsilon optoNCDT 2300) is used to track the position of the impactor head along the $z$-axis during impact and rebound, while  a force sensor is used to measure the impact force. The falling height is set such that the speed at the onset of impact in all cases is 1.85 m/s. The obtained data (impact force, slider position and time), along with the sample deflection, are collected via an acquisition system with a sampling rate of 20 kHz. The data is used to generate the force-time ($F-t$), deflection-time $(\delta-t$), and force-deflection ($F-\delta$) curves (See Supplementary Fig. 3b, 3c and 3d).

\backmatter

\bmhead*{Acknowledgments}
This research was carried out using the Core Technology Platform resources at New York University Abu Dhabi. In particular, we would like to acknowledge the support of the Advanced Manufacturing Core.

\section*{Declarations}

\begin{itemize}
\item \textbf{Supplementary Information} are provided as separate files with the submission.
\item \textbf{Competing interests:} The authors declare no competing interests.
\item \textbf{Data availability:} Data supporting the findings of this study are available from the corresponding author upon request.
\end{itemize}

\bibliography{References}

\begin{thebibliography}{10}
\expandafter\ifx\csname url\endcsname\relax
  \def\url#1{\burl{#1}}\fi
\expandafter\ifx\csname urlprefix\endcsname\relax\def\urlprefix{URL }\fi
\providecommand{\bibinfo}[2]{#2}
\providecommand{\eprint}[2][]{\url{#2}}
\providecommand{\doi}[1]{\url{https://doi.org/#1}}
\bibcommenthead

\bibitem{clapham_megaptera_1999}
\bibinfo{author}{Clapham, P.~J.} \& \bibinfo{author}{Mead, J.~G.}
\newblock \bibinfo{title}{Megaptera novaeangliae}.
\newblock \emph{\bibinfo{journal}{Mammalian Species}} ~(604),
  \bibinfo{pages}{1--9} (\bibinfo{year}{1999}).
\newblock \urlprefix\url{https://www.jstor.org/stable/3504352}.
\newblock \doi{10.2307/3504352}, \bibinfo{note}{publisher: [American Society of
  Mammalogists, Oxford University Press]} .

\bibitem{feng_super-hydrophobic_2002}
\bibinfo{author}{Feng, L.} \emph{et~al.}
\newblock \bibinfo{title}{Super-{Hydrophobic} {Surfaces}: {From} {Natural} to
  {Artificial}}.
\newblock \emph{\bibinfo{journal}{Advanced Materials}}
  \textbf{\bibinfo{volume}{14}}~(24), \bibinfo{pages}{1857--1860}
  (\bibinfo{year}{2002}).
\newblock
  \urlprefix\url{https://onlinelibrary.wiley.com/doi/abs/10.1002/adma.200290020}.
\newblock \doi{10.1002/adma.200290020}, \bibinfo{note}{\_eprint:
  https://onlinelibrary.wiley.com/doi/pdf/10.1002/adma.200290020} .

\bibitem{ha_review_2020}
\bibinfo{author}{Ha, N.~S.} \& \bibinfo{author}{Lu, G.}
\newblock \bibinfo{title}{A review of recent research on bio-inspired
  structures and materials for energy absorption applications}.
\newblock \emph{\bibinfo{journal}{Composites Part B: Engineering}}
  \textbf{\bibinfo{volume}{181}}, \bibinfo{pages}{107496}
  (\bibinfo{year}{2020}).
\newblock
  \urlprefix\url{https://www.sciencedirect.com/science/article/pii/S1359836819339964}.
\newblock \doi{10.1016/j.compositesb.2019.107496} .

\bibitem{weissengruber_structure_2006}
\bibinfo{author}{Weissengruber, G.~E.} \emph{et~al.}
\newblock \bibinfo{title}{The structure of the cushions in the feet of
  {African} elephants ({Loxodonta} africana)}.
\newblock \emph{\bibinfo{journal}{Journal of Anatomy}}
  \textbf{\bibinfo{volume}{209}}~(6), \bibinfo{pages}{781--792}
  (\bibinfo{year}{2006}).
\newblock
  \urlprefix\url{https://www.ncbi.nlm.nih.gov/pmc/articles/PMC2048995/}.
\newblock \doi{10.1111/j.1469-7580.2006.00648.x} .

\bibitem{wang_why_2011}
\bibinfo{author}{Wang, L.} \emph{et~al.}
\newblock \bibinfo{title}{Why {Do} {Woodpeckers} {Resist} {Head} {Impact}
  {Injury}: {A} {Biomechanical} {Investigation}}.
\newblock \emph{\bibinfo{journal}{PLOS ONE}} \textbf{\bibinfo{volume}{6}}~(10),
  \bibinfo{pages}{e26490} (\bibinfo{year}{2011}).
\newblock
  \urlprefix\url{https://journals.plos.org/plosone/article?id=10.1371/journal.pone.0026490}.
\newblock \doi{10.1371/journal.pone.0026490}, \bibinfo{note}{publisher: Public
  Library of Science} .

\bibitem{gibson_mechanics_1982}
\bibinfo{author}{Gibson, I.~J.} \& \bibinfo{author}{Ashby, M.~F.}
\newblock \bibinfo{title}{The mechanics of three-dimensional cellular
  materials}.
\newblock \emph{\bibinfo{journal}{Proceedings of the Royal Society of London.
  A. Mathematical and Physical Sciences}}
  \textbf{\bibinfo{volume}{382}}~(1782), \bibinfo{pages}{43--59}
  (\bibinfo{year}{1982}).
\newblock
  \urlprefix\url{https://royalsocietypublishing.org/doi/abs/10.1098/rspa.1982.0088}.
\newblock \doi{10.1098/rspa.1982.0088}, \bibinfo{note}{publisher: Royal
  Society} .

\bibitem{aubert_low-density_1985}
\bibinfo{author}{Aubert, J.~H.} \& \bibinfo{author}{Clough, R.~L.}
\newblock \bibinfo{title}{Low-density, microcellular polystyrene foams}.
\newblock \emph{\bibinfo{journal}{Polymer}} \textbf{\bibinfo{volume}{26}}~(13),
  \bibinfo{pages}{2047--2054} (\bibinfo{year}{1985}).
\newblock
  \urlprefix\url{https://www.sciencedirect.com/science/article/pii/0032386185901867}.
\newblock \doi{10.1016/0032-3861(85)90186-7} .

\bibitem{brumfield_characterization_1969}
\bibinfo{author}{Brumfield, H.} \& \bibinfo{author}{Estill, W.}
\newblock \bibinfo{title}{Characterization of {Foam} {Structure} by {Use} of
  the {Scanning} {Electron} {Microscope}}.
\newblock \emph{\bibinfo{journal}{Journal of Cellular Plastics}}
  \textbf{\bibinfo{volume}{5}}~(4), \bibinfo{pages}{212--220}
  (\bibinfo{year}{1969}).
\newblock \urlprefix\url{https://doi.org/10.1177/0021955X6900500403}.
\newblock \doi{10.1177/0021955X6900500403}, \bibinfo{note}{publisher: SAGE
  Publications Ltd STM} .

\bibitem{almanza_microestructure_2001}
\bibinfo{author}{Almanza, O.}, \bibinfo{author}{Rodriguez-Pérez, M.~A.} \&
  \bibinfo{author}{de~Saja, J.~A.}
\newblock \bibinfo{title}{The microestructure of polyethylene foams produced by
  a nitrogen solution process}.
\newblock \emph{\bibinfo{journal}{Polymer}} \textbf{\bibinfo{volume}{42}}~(16),
  \bibinfo{pages}{7117--7126} (\bibinfo{year}{2001}).
\newblock
  \urlprefix\url{https://www.sciencedirect.com/science/article/pii/S0032386101001070}.
\newblock \doi{10.1016/S0032-3861(01)00107-0} .

\bibitem{gibson_cellular_1999}
\bibinfo{author}{Gibson, L.~J.} \& \bibinfo{author}{Ashby, M.~F.}
\newblock \emph{\bibinfo{title}{Cellular {Solids}: {Structure} and
  {Properties}}} \bibinfo{edition}{2 edition} edn
  (\bibinfo{publisher}{Cambridge University Press},
  \bibinfo{address}{Cambridge}, \bibinfo{year}{1999}).

\bibitem{xiang_energy_2020}
\bibinfo{author}{Xiang, X.~M.}, \bibinfo{author}{Lu, G.} \&
  \bibinfo{author}{You, Z.}
\newblock \bibinfo{title}{Energy absorption of origami inspired structures and
  materials}.
\newblock \emph{\bibinfo{journal}{Thin-Walled Structures}}
  \textbf{\bibinfo{volume}{157}}, \bibinfo{pages}{107130}
  (\bibinfo{year}{2020}).
\newblock
  \urlprefix\url{https://www.sciencedirect.com/science/article/pii/S026382312031003X}.
\newblock \doi{10.1016/j.tws.2020.107130} .

\bibitem{wang_architected_2019}
\bibinfo{author}{Wang, Y.} \emph{et~al.}
\newblock \bibinfo{title}{Architected lattices with adaptive energy
  absorption}.
\newblock \emph{\bibinfo{journal}{Extreme Mechanics Letters}}
  \textbf{\bibinfo{volume}{33}}, \bibinfo{pages}{100557}
  (\bibinfo{year}{2019}).
\newblock
  \urlprefix\url{https://www.sciencedirect.com/science/article/pii/S2352431619302251}.
\newblock \doi{10.1016/j.eml.2019.100557} .

\bibitem{kucewicz_modelling_2018}
\bibinfo{author}{Kucewicz, M.}, \bibinfo{author}{Baranowski, P.},
  \bibinfo{author}{Małachowski, J.}, \bibinfo{author}{Popławski, A.} \&
  \bibinfo{author}{Płatek, P.}
\newblock \bibinfo{title}{Modelling, and characterization of {3D} printed
  cellular structures}.
\newblock \emph{\bibinfo{journal}{Materials \& Design}}
  \textbf{\bibinfo{volume}{142}}, \bibinfo{pages}{177--189}
  (\bibinfo{year}{2018}).
\newblock
  \urlprefix\url{https://www.sciencedirect.com/science/article/pii/S0264127518300364}.
\newblock \doi{10.1016/j.matdes.2018.01.028} .

\bibitem{mohsenizadeh_additively-manufactured_2018}
\bibinfo{author}{Mohsenizadeh, M.}, \bibinfo{author}{Gasbarri, F.},
  \bibinfo{author}{Munther, M.}, \bibinfo{author}{Beheshti, A.} \&
  \bibinfo{author}{Davami, K.}
\newblock \bibinfo{title}{Additively-manufactured lightweight {Metamaterials}
  for energy absorption}.
\newblock \emph{\bibinfo{journal}{Materials \& Design}}
  \textbf{\bibinfo{volume}{139}}, \bibinfo{pages}{521--530}
  (\bibinfo{year}{2018}).
\newblock
  \urlprefix\url{https://www.sciencedirect.com/science/article/pii/S0264127517310687}.
\newblock \doi{10.1016/j.matdes.2017.11.037} .

\bibitem{al-ketan_nature-inspired_2018}
\bibinfo{author}{Al-Ketan, O.}, \bibinfo{author}{Soliman, A.},
  \bibinfo{author}{AlQubaisi, A.~M.} \& \bibinfo{author}{Abu Al-Rub, R.~K.}
\newblock \bibinfo{title}{Nature-{Inspired} {Lightweight} {Cellular}
  {Co}-{Continuous} {Composites} with {Architected} {Periodic} {Gyroidal}
  {Structures}}.
\newblock \emph{\bibinfo{journal}{Advanced Engineering Materials}}
  \textbf{\bibinfo{volume}{20}}~(2), \bibinfo{pages}{1700549}
  (\bibinfo{year}{2018}).
\newblock
  \urlprefix\url{https://onlinelibrary.wiley.com/doi/abs/10.1002/adem.201700549}.
\newblock \doi{10.1002/adem.201700549}, \bibinfo{note}{\_eprint:
  https://onlinelibrary.wiley.com/doi/pdf/10.1002/adem.201700549} .

\bibitem{ashby_mechanical_1983}
\bibinfo{author}{Ashby, M.~F.} \& \bibinfo{author}{Medalist, R. F.~M.}
\newblock \bibinfo{title}{The mechanical properties of cellular solids}.
\newblock \emph{\bibinfo{journal}{Metallurgical Transactions A}}
  \textbf{\bibinfo{volume}{14}}~(9), \bibinfo{pages}{1755--1769}
  (\bibinfo{year}{1983}).
\newblock \urlprefix\url{https://doi.org/10.1007/BF02645546}.
\newblock \doi{10.1007/BF02645546} .

\bibitem{lakes_materials_1993}
\bibinfo{author}{Lakes, R.}
\newblock \bibinfo{title}{Materials with structural hierarchy}.
\newblock \emph{\bibinfo{journal}{Nature}}
  \textbf{\bibinfo{volume}{361}}~(6412), \bibinfo{pages}{511--515}
  (\bibinfo{year}{1993}).
\newblock \urlprefix\url{https://www.nature.com/articles/361511a0}.
\newblock \doi{10.1038/361511a0}, \bibinfo{note}{number: 6412 Publisher: Nature
  Publishing Group} .

\bibitem{gosselin_cell_2005}
\bibinfo{author}{Gosselin, R.} \& \bibinfo{author}{Rodrigue, D.}
\newblock \bibinfo{title}{Cell morphology analysis of high density polymer
  foams}.
\newblock \emph{\bibinfo{journal}{Polymer Testing}}
  \textbf{\bibinfo{volume}{24}}~(8), \bibinfo{pages}{1027--1035}
  (\bibinfo{year}{2005}).
\newblock
  \urlprefix\url{https://www.sciencedirect.com/science/article/pii/S014294180500108X}.
\newblock \doi{10.1016/j.polymertesting.2005.07.005} .

\bibitem{gibson_biomechanics_2005}
\bibinfo{author}{Gibson, L.~J.}
\newblock \bibinfo{title}{Biomechanics of cellular solids}.
\newblock \emph{\bibinfo{journal}{Journal of Biomechanics}}
  \textbf{\bibinfo{volume}{38}}~(3), \bibinfo{pages}{377--399}
  (\bibinfo{year}{2005}).
\newblock
  \urlprefix\url{https://www.sciencedirect.com/science/article/pii/S0021929004004919}.
\newblock \doi{10.1016/j.jbiomech.2004.09.027} .

\bibitem{schaedler_ultralight_2011}
\bibinfo{author}{Schaedler, T.~A.} \emph{et~al.}
\newblock \bibinfo{title}{Ultralight {Metallic} {Microlattices}}.
\newblock \emph{\bibinfo{journal}{Science}}
  \textbf{\bibinfo{volume}{334}}~(6058), \bibinfo{pages}{962--965}
  (\bibinfo{year}{2011}).
\newblock
  \urlprefix\url{https://www.science.org/doi/full/10.1126/science.1211649}.
\newblock \doi{10.1126/science.1211649}, \bibinfo{note}{publisher: American
  Association for the Advancement of Science} .

\bibitem{kuang_advances_2019}
\bibinfo{author}{Kuang, X.} \emph{et~al.}
\newblock \bibinfo{title}{Advances in {4D} {Printing}: {Materials} and
  {Applications}}.
\newblock \emph{\bibinfo{journal}{Advanced Functional Materials}}
  \textbf{\bibinfo{volume}{29}}~(2), \bibinfo{pages}{1805290}
  (\bibinfo{year}{2019}).
\newblock
  \urlprefix\url{https://onlinelibrary.wiley.com/doi/abs/10.1002/adfm.201805290}.
\newblock \doi{10.1002/adfm.201805290}, \bibinfo{note}{\_eprint:
  https://onlinelibrary.wiley.com/doi/pdf/10.1002/adfm.201805290} .

\bibitem{thakar_3d_2022}
\bibinfo{author}{Thakar, C.~M.} \emph{et~al.}
\newblock \bibinfo{title}{3d {Printing}: {Basic} principles and applications}.
\newblock \emph{\bibinfo{journal}{Materials Today: Proceedings}}
  \textbf{\bibinfo{volume}{51}}, \bibinfo{pages}{842--849}
  (\bibinfo{year}{2022}).
\newblock
  \urlprefix\url{https://www.sciencedirect.com/science/article/pii/S2214785321046575}.
\newblock \doi{10.1016/j.matpr.2021.06.272} .

\bibitem{teunis_4d_2021}
\bibinfo{author}{Teunis, v.~M.}, \bibinfo{author}{Link to~external site, t. l.
  w. o. i. a. n.~w.}, \bibinfo{author}{Shahram, J.}, \bibinfo{author}{B, J.
  K.~M.} \& \bibinfo{author}{Zadpoor, A.~A.}
\newblock \bibinfo{title}{{4D} printing of reconfigurable metamaterials and
  devices}.
\newblock \emph{\bibinfo{journal}{Communications Materials}}
  \textbf{\bibinfo{volume}{2}}~(1) (\bibinfo{year}{2021}).
\newblock
  \urlprefix\url{https://www.proquest.com/docview/2536652098/abstract/4CA579280C4A4C09PQ/1}.
\newblock \doi{https://doi.org/10.1038/s43246-021-00165-8},
  \bibinfo{note}{place: London, United States Publisher: Nature Publishing
  Group} .

\bibitem{dalaq_finite_2016}
\bibinfo{author}{Dalaq, A.~S.}, \bibinfo{author}{Abueidda, D.~W.},
  \bibinfo{author}{Abu Al-Rub, R.~K.} \& \bibinfo{author}{Jasiuk, I.~M.}
\newblock \bibinfo{title}{Finite element prediction of effective elastic
  properties of interpenetrating phase composites with architectured {3D} sheet
  reinforcements}.
\newblock \emph{\bibinfo{journal}{International Journal of Solids and
  Structures}} \textbf{\bibinfo{volume}{83}}, \bibinfo{pages}{169--182}
  (\bibinfo{year}{2016}).
\newblock
  \urlprefix\url{https://www.sciencedirect.com/science/article/pii/S0020768316000202}.
\newblock \doi{10.1016/j.ijsolstr.2016.01.011} .

\bibitem{han_microscopic_2017}
\bibinfo{author}{Han, S.~C.}, \bibinfo{author}{Choi, J.~M.},
  \bibinfo{author}{Liu, G.} \& \bibinfo{author}{Kang, K.}
\newblock \bibinfo{title}{A {Microscopic} {Shell} {Structure} with
  {Schwarz}’s {D}-{Surface}}.
\newblock \emph{\bibinfo{journal}{Scientific Reports}}
  \textbf{\bibinfo{volume}{7}}~(1), \bibinfo{pages}{13405}
  (\bibinfo{year}{2017}).
\newblock \urlprefix\url{https://www.nature.com/articles/s41598-017-13618-3}.
\newblock \doi{10.1038/s41598-017-13618-3}, \bibinfo{note}{number: 1 Publisher:
  Nature Publishing Group} .

\bibitem{jiang_highly-stretchable_2016}
\bibinfo{author}{Jiang, Y.} \& \bibinfo{author}{Wang, Q.}
\newblock \bibinfo{title}{Highly-stretchable {3D}-architected {Mechanical}
  {Metamaterials}}.
\newblock \emph{\bibinfo{journal}{Scientific Reports}}
  \textbf{\bibinfo{volume}{6}}~(1), \bibinfo{pages}{34147}
  (\bibinfo{year}{2016}).
\newblock \urlprefix\url{https://www.nature.com/articles/srep34147}.
\newblock \doi{10.1038/srep34147}, \bibinfo{note}{number: 1 Publisher: Nature
  Publishing Group} .

\bibitem{dalaq_mechanical_2016}
\bibinfo{author}{Dalaq, A.~S.}, \bibinfo{author}{Abueidda, D.~W.} \&
  \bibinfo{author}{Abu Al-Rub, R.~K.}
\newblock \bibinfo{title}{Mechanical properties of {3D} printed
  interpenetrating phase composites with novel architectured {3D} solid-sheet
  reinforcements}.
\newblock \emph{\bibinfo{journal}{Composites Part A: Applied Science and
  Manufacturing}} \textbf{\bibinfo{volume}{84}}, \bibinfo{pages}{266--280}
  (\bibinfo{year}{2016}).
\newblock
  \urlprefix\url{https://www.sciencedirect.com/science/article/pii/S1359835X1600066X}.
\newblock \doi{10.1016/j.compositesa.2016.02.009} .

\bibitem{abou-ali_mechanical_2020}
\bibinfo{author}{Abou-Ali, A.~M.}, \bibinfo{author}{Al-Ketan, O.},
  \bibinfo{author}{Lee, D.-W.}, \bibinfo{author}{Rowshan, R.} \&
  \bibinfo{author}{Abu Al-Rub, R.~K.}
\newblock \bibinfo{title}{Mechanical behavior of polymeric selective laser
  sintered ligament and sheet based lattices of triply periodic minimal surface
  architectures}.
\newblock \emph{\bibinfo{journal}{Materials \& Design}}
  \textbf{\bibinfo{volume}{196}}, \bibinfo{pages}{109100}
  (\bibinfo{year}{2020}).
\newblock
  \urlprefix\url{https://www.sciencedirect.com/science/article/pii/S0264127520306353}.
\newblock \doi{10.1016/j.matdes.2020.109100} .

\bibitem{habib_fabrication_2018}
\bibinfo{author}{Habib, F.~N.}, \bibinfo{author}{Iovenitti, P.},
  \bibinfo{author}{Masood, S.~H.} \& \bibinfo{author}{Nikzad, M.}
\newblock \bibinfo{title}{Fabrication of polymeric lattice structures for
  optimum energy absorption using {Multi} {Jet} {Fusion} technology}.
\newblock \emph{\bibinfo{journal}{Materials \& Design}}
  \textbf{\bibinfo{volume}{155}}, \bibinfo{pages}{86--98}
  (\bibinfo{year}{2018}).
\newblock
  \urlprefix\url{https://www.sciencedirect.com/science/article/pii/S0264127518304441}.
\newblock \doi{10.1016/j.matdes.2018.05.059} .

\bibitem{li_additive_2021-1}
\bibinfo{author}{Li, T.}, \bibinfo{author}{Jarrar, F.}, \bibinfo{author}{Abu
  Al-Rub, R.} \& \bibinfo{author}{Cantwell, W.}
\newblock \bibinfo{title}{Additive manufactured semi-plate lattice materials
  with high stiffness, strength and toughness}.
\newblock \emph{\bibinfo{journal}{International Journal of Solids and
  Structures}} \textbf{\bibinfo{volume}{230-231}}, \bibinfo{pages}{111153}
  (\bibinfo{year}{2021}).
\newblock
  \urlprefix\url{https://www.sciencedirect.com/science/article/pii/S0020768321002432}.
\newblock \doi{10.1016/j.ijsolstr.2021.111153} .

\bibitem{abueidda_effective_2016}
\bibinfo{author}{Abueidda, D.~W.} \emph{et~al.}
\newblock \bibinfo{title}{Effective conductivities and elastic moduli of novel
  foams with triply periodic minimal surfaces}.
\newblock \emph{\bibinfo{journal}{Mechanics of Materials}}
  \textbf{\bibinfo{volume}{95}}, \bibinfo{pages}{102--115}
  (\bibinfo{year}{2016}).
\newblock
  \urlprefix\url{https://www.sciencedirect.com/science/article/pii/S0167663616000053}.
\newblock \doi{10.1016/j.mechmat.2016.01.004} .

\bibitem{qureshi_effect_2022}
\bibinfo{author}{Qureshi, Z.~A.}, \bibinfo{author}{Addin Burhan Al-Omari, S.},
  \bibinfo{author}{Elnajjar, E.}, \bibinfo{author}{Al-Ketan, O.} \&
  \bibinfo{author}{Al-Rub, R.~A.}
\newblock \bibinfo{title}{On the effect of porosity and functional grading of
  {3D} printable triply periodic minimal surface ({TPMS}) based architected
  lattices embedded with a phase change material}.
\newblock \emph{\bibinfo{journal}{International Journal of Heat and Mass
  Transfer}} \textbf{\bibinfo{volume}{183}}, \bibinfo{pages}{122111}
  (\bibinfo{year}{2022}).
\newblock
  \urlprefix\url{https://www.sciencedirect.com/science/article/pii/S0017931021012175}.
\newblock \doi{10.1016/j.ijheatmasstransfer.2021.122111} .

\bibitem{abueidda_micromechanical_2015}
\bibinfo{author}{Abueidda, D.~W.}, \bibinfo{author}{Dalaq, A.~S.},
  \bibinfo{author}{Abu Al-Rub, R.~K.} \& \bibinfo{author}{Jasiuk, I.}
\newblock \bibinfo{title}{Micromechanical finite element predictions of a
  reduced coefficient of thermal expansion for {3D} periodic architectured
  interpenetrating phase composites}.
\newblock \emph{\bibinfo{journal}{Composite Structures}}
  \textbf{\bibinfo{volume}{133}}, \bibinfo{pages}{85--97}
  (\bibinfo{year}{2015}).
\newblock
  \urlprefix\url{https://www.sciencedirect.com/science/article/pii/S0263822315005991}.
\newblock \doi{10.1016/j.compstruct.2015.06.082} .

\bibitem{baobaid_fluid_2022}
\bibinfo{author}{Baobaid, N.}, \bibinfo{author}{Ali, M.~I.},
  \bibinfo{author}{Khan, K.~A.} \& \bibinfo{author}{Abu Al-Rub, R.~K.}
\newblock \bibinfo{title}{Fluid flow and heat transfer of porous {TPMS}
  architected heat sinks in free convection environment}.
\newblock \emph{\bibinfo{journal}{Case Studies in Thermal Engineering}}
  \textbf{\bibinfo{volume}{33}}, \bibinfo{pages}{101944}
  (\bibinfo{year}{2022}).
\newblock
  \urlprefix\url{https://www.sciencedirect.com/science/article/pii/S2214157X22001903}.
\newblock \doi{10.1016/j.csite.2022.101944} .

\bibitem{alqahtani_thermal_2021}
\bibinfo{author}{Alqahtani, S.}, \bibinfo{author}{Ali, H.~M.},
  \bibinfo{author}{Farukh, F.}, \bibinfo{author}{Silberschmidt, V.~V.} \&
  \bibinfo{author}{Kandan, K.}
\newblock \bibinfo{title}{Thermal performance of additively manufactured
  polymer lattices}.
\newblock \emph{\bibinfo{journal}{Journal of Building Engineering}}
  \textbf{\bibinfo{volume}{39}}, \bibinfo{pages}{102243}
  (\bibinfo{year}{2021}).
\newblock
  \urlprefix\url{https://www.sciencedirect.com/science/article/pii/S2352710221000991}.
\newblock \doi{10.1016/j.jobe.2021.102243} .

\bibitem{abueidda_compression_2020}
\bibinfo{author}{Abueidda, D.~W.}, \bibinfo{author}{Elhebeary, M.},
  \bibinfo{author}{Shiang, C.-S.~A.}, \bibinfo{author}{Abu Al-Rub, R.~K.} \&
  \bibinfo{author}{Jasiuk, I.~M.}
\newblock \bibinfo{title}{Compression and buckling of microarchitectured
  {Neovius}-lattice}.
\newblock \emph{\bibinfo{journal}{Extreme Mechanics Letters}}
  \textbf{\bibinfo{volume}{37}}, \bibinfo{pages}{100688}
  (\bibinfo{year}{2020}).
\newblock
  \urlprefix\url{https://www.sciencedirect.com/science/article/pii/S2352431620300596}.
\newblock \doi{10.1016/j.eml.2020.100688} .

\bibitem{song_octet-truss_2019}
\bibinfo{author}{Song, J.} \emph{et~al.}
\newblock \bibinfo{title}{Octet-truss cellular materials for improved
  mechanical properties and specific energy absorption}.
\newblock \emph{\bibinfo{journal}{Materials \& Design}}
  \textbf{\bibinfo{volume}{173}}, \bibinfo{pages}{107773}
  (\bibinfo{year}{2019}).
\newblock
  \urlprefix\url{https://www.sciencedirect.com/science/article/pii/S0264127519302102}.
\newblock \doi{10.1016/j.matdes.2019.107773} .

\bibitem{lakes_viscoelastic_1998}
\bibinfo{author}{Lakes, R.~S.}
\newblock \emph{\bibinfo{title}{Viscoelastic {Solids}}}
  (\bibinfo{publisher}{CRC Press}, \bibinfo{year}{1998}).
\newblock \bibinfo{note}{Google-Books-ID: soZZl17sm5IC}.

\bibitem{ramirez_viscoelastic_2018}
\bibinfo{author}{Ramirez, B.~J.}, \bibinfo{author}{Misra, U.} \&
  \bibinfo{author}{Gupta, V.}
\newblock \bibinfo{title}{Viscoelastic foam-filled lattice for high energy
  absorption}.
\newblock \emph{\bibinfo{journal}{Mechanics of Materials}}
  \textbf{\bibinfo{volume}{127}}, \bibinfo{pages}{39--47}
  (\bibinfo{year}{2018}).
\newblock
  \urlprefix\url{https://www.sciencedirect.com/science/article/pii/S0167663617308293}.
\newblock \doi{10.1016/j.mechmat.2018.08.011} .

\bibitem{lang_science_2007}
\bibinfo{author}{Lang, R.~J.}
\newblock \bibinfo{title}{The science of origami}.
\newblock \emph{\bibinfo{journal}{Physics World}}
  \textbf{\bibinfo{volume}{20}}~(2), \bibinfo{pages}{30--31}
  (\bibinfo{year}{2007}).
\newblock \urlprefix\url{https://doi.org/10.1088/2058-7058/20/2/31}.
\newblock \doi{10.1088/2058-7058/20/2/31}, \bibinfo{note}{publisher: IOP
  Publishing} .

\bibitem{kresling_fifth_2020}
\bibinfo{author}{Kresling, B.}
\newblock \bibinfo{title}{The {Fifth} {Fold}: {Complex} {Symmetries} in
  {Kresling}-origami {Patterns}}.
\newblock \emph{\bibinfo{journal}{Symmetry: Culture and Science}}
  \textbf{\bibinfo{volume}{31}}, \bibinfo{pages}{403--416}
  (\bibinfo{year}{2020}).
\newblock \doi{10.26830/symmetry_2020_4_403} .

\bibitem{dalaq_experimentally-validated_2022}
\bibinfo{author}{Dalaq, A.~S.} \& \bibinfo{author}{Daqaq, M.~F.}
\newblock \bibinfo{title}{Experimentally-validated computational modeling and
  characterization of the quasi-static behavior of functional {3D}-printed
  origami-inspired springs}.
\newblock \emph{\bibinfo{journal}{Materials \& Design}}
  \textbf{\bibinfo{volume}{216}}, \bibinfo{pages}{110541}
  (\bibinfo{year}{2022}).
\newblock
  \urlprefix\url{https://www.sciencedirect.com/science/article/pii/S0264127522001629}.
\newblock \doi{10.1016/j.matdes.2022.110541} .

\bibitem{kresling_origami-structures_2012}
\bibinfo{author}{Kresling, B.}
\newblock \bibinfo{title}{Origami-structures in nature: lessons in designing
  “smart” materials}.
\newblock \emph{\bibinfo{journal}{MRS Online Proceedings Library (OPL)}}
  \textbf{\bibinfo{volume}{1420}} (\bibinfo{year}{2012}).
\newblock
  \urlprefix\url{https://www.cambridge.org/core/journals/mrs-online-proceedings-library-archive/article/abs/origamistructures-in-nature-lessons-in-designing-smart-materials/46181BA89E24CE3EFB598B411CBA9D96}.
\newblock \doi{10.1557/opl.2012.536}, \bibinfo{note}{publisher: Cambridge
  University Press} .

\bibitem{khazaaleh_combining_2022}
\bibinfo{author}{Khazaaleh, S.}, \bibinfo{author}{Masana, R.} \&
  \bibinfo{author}{Daqaq, M.~F.}
\newblock \bibinfo{title}{Combining advanced {3D} printing technologies with
  origami principles: {A} new paradigm for the design of functional, durable,
  and scalable springs}.
\newblock \emph{\bibinfo{journal}{Composites Part B: Engineering}}
  \textbf{\bibinfo{volume}{236}}, \bibinfo{pages}{109811}
  (\bibinfo{year}{2022}).
\newblock
  \urlprefix\url{https://www.sciencedirect.com/science/article/pii/S1359836822001913}.
\newblock \doi{10.1016/j.compositesb.2022.109811} .

\bibitem{zhai_origami-inspired_2018}
\bibinfo{author}{Zhai, Z.}, \bibinfo{author}{Wang, Y.} \&
  \bibinfo{author}{Jiang, H.}
\newblock \bibinfo{title}{Origami-inspired, on-demand deployable and
  collapsible mechanical metamaterials with tunable stiffness}.
\newblock \emph{\bibinfo{journal}{Proceedings of the National Academy of
  Sciences}} \textbf{\bibinfo{volume}{115}}~(9), \bibinfo{pages}{2032--2037}
  (\bibinfo{year}{2018}).
\newblock \urlprefix\url{https://www.pnas.org/doi/10.1073/pnas.1720171115}.
\newblock \doi{10.1073/pnas.1720171115}, \bibinfo{note}{publisher: Proceedings
  of the National Academy of Sciences} .

\bibitem{hunt_twist_2005}
\bibinfo{author}{Hunt, G.~W.} \& \bibinfo{author}{Ario, I.}
\newblock \bibinfo{title}{Twist buckling and the foldable cylinder: an exercise
  in origami}.
\newblock \emph{\bibinfo{journal}{International Journal of Non-Linear
  Mechanics}} \textbf{\bibinfo{volume}{40}}~(6), \bibinfo{pages}{833--843}
  (\bibinfo{year}{2005}).
\newblock
  \urlprefix\url{https://www.sciencedirect.com/science/article/pii/S0020746204001581}.
\newblock \doi{10.1016/j.ijnonlinmec.2004.08.011} .

\bibitem{noauthor_kresling-pattern_2017}
\bibinfo{title}{The {Kresling}-{Pattern} and our origami world}
  (\bibinfo{year}{2017}).
\newblock
  \urlprefix\url{https://thekidshouldseethis.com/post/the-kresling-pattern-and-our-origami-world}.

\bibitem{li_compressive_2006}
\bibinfo{author}{Li, Q.~M.}, \bibinfo{author}{Magkiriadis, I.} \&
  \bibinfo{author}{Harrigan, J.~J.}
\newblock \bibinfo{title}{Compressive {Strain} at the {Onset} of
  {Densification} of {Cellular} {Solids}}.
\newblock \emph{\bibinfo{journal}{Journal of Cellular Plastics}}
  \textbf{\bibinfo{volume}{42}}~(5), \bibinfo{pages}{371--392}
  (\bibinfo{year}{2006}).
\newblock \urlprefix\url{https://doi.org/10.1177/0021955X06063519}.
\newblock \doi{10.1177/0021955X06063519}, \bibinfo{note}{publisher: SAGE
  Publications Ltd STM} .

\bibitem{storakers_material_1986}
\bibinfo{author}{Storåkers, B.}
\newblock \bibinfo{title}{On material representation and constitutive branching
  in finite compressible elasticity}.
\newblock \emph{\bibinfo{journal}{Journal of the Mechanics and Physics of
  Solids}} \textbf{\bibinfo{volume}{34}}~(2), \bibinfo{pages}{125--145}
  (\bibinfo{year}{1986}).
\newblock
  \urlprefix\url{https://www.sciencedirect.com/science/article/pii/0022509686900335}.
\newblock \doi{10.1016/0022-5096(86)90033-5} .

\bibitem{ishida_design_2017}
\bibinfo{author}{Ishida, S.}, \bibinfo{author}{Suzuki, K.} \&
  \bibinfo{author}{Shimosaka, H.}
\newblock \bibinfo{title}{Design and {Experimental} {Analysis} of
  {Origami}-{Inspired} {Vibration} {Isolator} {With} {Quasi}-{Zero}-{Stiffness}
  {Characteristic}}.
\newblock \emph{\bibinfo{journal}{Journal of Vibration and Acoustics}}
  \textbf{\bibinfo{volume}{139}}~(5) (\bibinfo{year}{2017}).
\newblock \urlprefix\url{https://doi.org/10.1115/1.4036465}.
\newblock \doi{10.1115/1.4036465} .

\bibitem{yasuda_origami-based_2017}
\bibinfo{author}{Yasuda, H.}, \bibinfo{author}{Tachi, T.},
  \bibinfo{author}{Lee, M.} \& \bibinfo{author}{Yang, J.}
\newblock \bibinfo{title}{Origami-based tunable truss structures for
  non-volatile mechanical memory operation}.
\newblock \emph{\bibinfo{journal}{Nature Communications}}
  \textbf{\bibinfo{volume}{8}}~(1), \bibinfo{pages}{962}
  (\bibinfo{year}{2017}).
\newblock \urlprefix\url{https://www.nature.com/articles/s41467-017-00670-w}.
\newblock \doi{10.1038/s41467-017-00670-w}, \bibinfo{note}{number: 1 Publisher:
  Nature Publishing Group} .

\bibitem{deleo_origami-based_2020}
\bibinfo{author}{Deleo, A.~A.}, \bibinfo{author}{O’Neil, J.},
  \bibinfo{author}{Yasuda, H.}, \bibinfo{author}{Salviato, M.} \&
  \bibinfo{author}{Yang, J.}
\newblock \bibinfo{title}{Origami-based deployable structures made of carbon
  fiber reinforced polymer composites}.
\newblock \emph{\bibinfo{journal}{Composites Science and Technology}}
  \textbf{\bibinfo{volume}{191}}, \bibinfo{pages}{108060}
  (\bibinfo{year}{2020}).
\newblock
  \urlprefix\url{https://www.sciencedirect.com/science/article/pii/S0266353819324996}.
\newblock \doi{10.1016/j.compscitech.2020.108060} .

\bibitem{butler_highly_2017}
\bibinfo{author}{Butler, J.}
\newblock \bibinfo{title}{Highly {Compressible} {Origami} {Bellows} for {Harsh}
  {Environments}}.
\newblock \emph{\bibinfo{journal}{Theses and Dissertations}}
  (\bibinfo{year}{2017}).
\newblock \urlprefix\url{https://scholarsarchive.byu.edu/etd/6657} .

\bibitem{liu_dynamic_2014}
\bibinfo{author}{Liu, Y.}, \bibinfo{author}{Schaedler, T.~A.} \&
  \bibinfo{author}{Chen, X.}
\newblock \bibinfo{title}{Dynamic energy absorption characteristics of hollow
  microlattice structures}.
\newblock \emph{\bibinfo{journal}{Mechanics of Materials}}
  \textbf{\bibinfo{volume}{77}}, \bibinfo{pages}{1--13} (\bibinfo{year}{2014}).
\newblock
  \urlprefix\url{https://www.sciencedirect.com/science/article/pii/S0167663614001070}.
\newblock \doi{10.1016/j.mechmat.2014.06.008} .

\bibitem{duocel_aluminum_2023}
\bibinfo{author}{Duocel}.
\newblock \bibinfo{title}{Aluminum foam}.
\newblock \urlprefix\url{https://www.duocelfoam.com/product-p/6010-028.htm}.

\bibitem{mieszala_micromechanics_2017}
\bibinfo{author}{Mieszala, M.} \emph{et~al.}
\newblock \bibinfo{title}{Micromechanics of {Amorphous} {Metal}/{Polymer}
  {Hybrid} {Structures} with {3D} {Cellular} {Architectures}: {Size} {Effects},
  {Buckling} {Behavior}, and {Energy} {Absorption} {Capability}}.
\newblock \emph{\bibinfo{journal}{Small}} \textbf{\bibinfo{volume}{13}}~(8),
  \bibinfo{pages}{1602514} (\bibinfo{year}{2017}).
\newblock
  \urlprefix\url{https://onlinelibrary.wiley.com/doi/abs/10.1002/smll.201602514}.
\newblock \doi{10.1002/smll.201602514}, \bibinfo{note}{\_eprint:
  https://onlinelibrary.wiley.com/doi/pdf/10.1002/smll.201602514} .

\bibitem{caserta_shock_2011}
\bibinfo{author}{Caserta, G.~D.}, \bibinfo{author}{Iannucci, L.} \&
  \bibinfo{author}{Galvanetto, U.}
\newblock \bibinfo{title}{Shock absorption performance of a motorbike helmet
  with honeycomb reinforced liner}.
\newblock \emph{\bibinfo{journal}{Composite Structures}}
  \textbf{\bibinfo{volume}{93}}~(11), \bibinfo{pages}{2748--2759}
  (\bibinfo{year}{2011}).
\newblock
  \urlprefix\url{https://www.sciencedirect.com/science/article/pii/S0263822311002133}.
\newblock \doi{10.1016/j.compstruct.2011.05.029} .

\bibitem{de_sousa_assessing_2012}
\bibinfo{author}{de~Sousa, R.~A.}, \bibinfo{author}{Gonçalves, D.},
  \bibinfo{author}{Coelho, R.} \& \bibinfo{author}{Teixeira-Dias, F.}
\newblock \bibinfo{title}{Assessing the effectiveness of a natural cellular
  material used as safety padding material in motorcycle helmets}.
\newblock \emph{\bibinfo{journal}{SIMULATION}}
  \textbf{\bibinfo{volume}{88}}~(5), \bibinfo{pages}{580--591}
  (\bibinfo{year}{2012}).
\newblock \urlprefix\url{https://doi.org/10.1177/0037549711414735}.
\newblock \doi{10.1177/0037549711414735}, \bibinfo{note}{publisher: SAGE
  Publications Ltd STM} .

\bibitem{zhu_parameterized_2014}
\bibinfo{author}{Hutchinson, J.~W.}, \bibinfo{author}{Mear, M.~E.} \&
  \bibinfo{author}{Rice, J.~R.}
\newblock \bibinfo{title}{{Crack Paralleling an Interface Between Dissimilar
  Materials}}.
\newblock \emph{\bibinfo{journal}{Journal of Applied Mechanics}}
  \textbf{\bibinfo{volume}{54}}~(4), \bibinfo{pages}{828--832}
  (\bibinfo{year}{1987}).
\newblock \urlprefix\url{https://doi.org/10.1115/1.3173124}.
\newblock \doi{10.1115/1.3173124},
  \bibinfo{eprint}{{\href{https://arxiv.org/abs/https://asmedigitalcollection.asme.org/appliedmechanics/article-pdf/54/4/828/5459510/828\_1.pdf}{{https://asmedigitalcollection.asme.org/appliedmechanics/article-pdf/54/4/828/5459510/828\_1.pdf}}}}
  .

\bibitem{sharma_bio-inspired_2022}
\bibinfo{author}{Sharma, D.} \& \bibinfo{author}{Hiremath, S.~S.}
\newblock \bibinfo{title}{Bio-inspired repeatable lattice structures for energy
  absorption: Experimental and finite element study}.
\newblock \emph{\bibinfo{journal}{Composite Structures}}
  \textbf{\bibinfo{volume}{283}}, \bibinfo{pages}{115102}
  (\bibinfo{year}{2022}) .

\bibitem{yang_advanced_2020}
\bibinfo{author}{Yang, C.} \& \bibinfo{author}{Li, Q.~M.}
\newblock \bibinfo{title}{Advanced lattice material with high energy absorption
  based on topology optimisation}.
\newblock \emph{\bibinfo{journal}{Mechanics of Materials}}
  \textbf{\bibinfo{volume}{148}}, \bibinfo{pages}{103536}
  (\bibinfo{year}{2020}).
\newblock
  \urlprefix\url{https://www.sciencedirect.com/science/article/pii/S0167663620305780}.
\newblock \doi{10.1016/j.mechmat.2020.103536} .

\bibitem{cadman_cuttlebone_2012}
\bibinfo{author}{Cadman, J.}, \bibinfo{author}{Zhou, S.},
  \bibinfo{author}{Chen, Y.} \& \bibinfo{author}{Li, Q.}
\newblock \bibinfo{title}{Cuttlebone: characterisation, application and
  development of biomimetic materials}.
\newblock \emph{\bibinfo{journal}{Journal of Bionic Engineering}}
  \textbf{\bibinfo{volume}{9}}~(3), \bibinfo{pages}{367--376}
  (\bibinfo{year}{2012}) .

\bibitem{stratasys_feel_2022}
\bibinfo{author}{Stratasys}.
\newblock \bibinfo{title}{Feel the difference. {Stratasys} {J750}\&trade;
  {Digital} {Anatomy}\&trade; {3D} {Printer}.} (\bibinfo{year}{2022}).
\newblock
  \urlprefix\url{https://www.stratasys.com/3d-printers/j750-digital-anatomy}.

\bibitem{stratasys_tango_2018}
\bibinfo{author}{Stratasys}.
\newblock \bibinfo{title}{Tango: {A} {Soft} {Flexible} {3D} {Printing}
  {Material}} (\bibinfo{year}{2018}).
\newblock
  \urlprefix\url{https://www.stratasys.com/en/materials/materials-catalog/polyjet-materials/tango/}.

\bibitem{object_fullcure_2009}
\bibinfo{author}{Object}.
\newblock \bibinfo{title}{{FullCure}® {Materials}} (\bibinfo{year}{2009}).
\newblock
  \urlprefix\url{http://svl.wpi.edu/wp-content/uploads/2014/04/FullCure_Letter_low-1.pdf}.

\bibitem{stratasys_vero_2018}
\bibinfo{author}{Stratasys}.
\newblock \bibinfo{title}{Vero: {A} {Realistic} {Multi}-{Color} {3D} {Printing}
  {Material} - {Stratasys}} (\bibinfo{year}{2018}).
\newblock
  \urlprefix\url{https://www.stratasys.com/en/materials/materials-catalog/polyjet-materials/vero/}.

\bibitem{stratasys_find_2022}
\bibinfo{author}{Stratasys}.
\newblock \bibinfo{title}{Find {Materials} and {Filaments} for {3D} {Printing}}
  (\bibinfo{year}{2022}).
\newblock \urlprefix\url{https://www.stratasys.com/materials/search}.

\bibitem{illinois_tool_works_inc_out_2020}
\bibinfo{author}{Inc, I. T.~W.}
\newblock \bibinfo{title}{Out of {Production} 5900 {Series} {Universal}
  {Testing} {Systems};} (\bibinfo{year}{2020}).
\newblock
  \urlprefix\url{http://www.instron.com/en/products/testing-systems/universal-testing-systems/low-force-universal-testing-systems/5900-series}.

\bibitem{simulia_abaqus_2022}
\bibinfo{author}{SIMULIA}.
\newblock \bibinfo{title}{Abaqus {Unified} {FEA} - {SIMULIA}™ by {Dassault}
  {Systèmes}®} (\bibinfo{year}{2022}).
\newblock
  \urlprefix\url{https://www.3ds.com/products-services/simulia/products/abaqus/}.

\bibitem{michael_smith_abaqusstandard_2019}
\bibinfo{author}{Smith, M.}
\newblock \emph{\bibinfo{title}{{ABAQUS}/{Standard} {User}'s {Manual}}}
  (\bibinfo{publisher}{Dassault Syst`emes Simulia Corp},
  \bibinfo{address}{United States}, \bibinfo{year}{2019}).

\bibitem{kim_introduction_2008}
\bibinfo{author}{Kim, N.-H.} \& \bibinfo{author}{Sankar, B.~V.}
\newblock \emph{\bibinfo{title}{Introduction to {Finite} {Element} {Analysis}
  and {Design}}} \bibinfo{edition}{1st edition} edn
  (\bibinfo{publisher}{Wiley}, \bibinfo{address}{New York Weinheim},
  \bibinfo{year}{2008}).

\end{thebibliography}



\begin{thebibliography}{6}
\ifx \bisbn   \undefined \def \bisbn  #1{ISBN #1}\fi
\ifx \binits  \undefined \def \binits#1{#1}\fi
\ifx \bauthor  \undefined \def \bauthor#1{#1}\fi
\ifx \batitle  \undefined \def \batitle#1{#1}\fi
\ifx \bjtitle  \undefined \def \bjtitle#1{#1}\fi
\ifx \bvolume  \undefined \def \bvolume#1{\textbf{#1}}\fi
\ifx \byear  \undefined \def \byear#1{#1}\fi
\ifx \bissue  \undefined \def \bissue#1{#1}\fi
\ifx \bfpage  \undefined \def \bfpage#1{#1}\fi
\ifx \blpage  \undefined \def \blpage #1{#1}\fi
\ifx \burl  \undefined \def \burl#1{\textsf{#1}}\fi
\ifx \doiurl  \undefined \def \doiurl#1{\url{https://doi.org/#1}}\fi
\ifx \betal  \undefined \def \betal{\textit{et al.}}\fi
\ifx \binstitute  \undefined \def \binstitute#1{#1}\fi
\ifx \binstitutionaled  \undefined \def \binstitutionaled#1{#1}\fi
\ifx \bctitle  \undefined \def \bctitle#1{#1}\fi
\ifx \beditor  \undefined \def \beditor#1{#1}\fi
\ifx \bpublisher  \undefined \def \bpublisher#1{#1}\fi
\ifx \bbtitle  \undefined \def \bbtitle#1{#1}\fi
\ifx \bedition  \undefined \def \bedition#1{#1}\fi
\ifx \bseriesno  \undefined \def \bseriesno#1{#1}\fi
\ifx \blocation  \undefined \def \blocation#1{#1}\fi
\ifx \bsertitle  \undefined \def \bsertitle#1{#1}\fi
\ifx \bsnm \undefined \def \bsnm#1{#1}\fi
\ifx \bsuffix \undefined \def \bsuffix#1{#1}\fi
\ifx \bparticle \undefined \def \bparticle#1{#1}\fi
\ifx \barticle \undefined \def \barticle#1{#1}\fi
\bibcommenthead
\ifx \bconfdate \undefined \def \bconfdate #1{#1}\fi
\ifx \botherref \undefined \def \botherref #1{#1}\fi
\ifx \url \undefined \def \url#1{\textsf{#1}}\fi
\ifx \bchapter \undefined \def \bchapter#1{#1}\fi
\ifx \bbook \undefined \def \bbook#1{#1}\fi
\ifx \bcomment \undefined \def \bcomment#1{#1}\fi
\ifx \oauthor \undefined \def \oauthor#1{#1}\fi
\ifx \citeauthoryear \undefined \def \citeauthoryear#1{#1}\fi
\ifx \endbibitem  \undefined \def \endbibitem {}\fi
\ifx \bconflocation  \undefined \def \bconflocation#1{#1}\fi
\ifx \arxivurl  \undefined \def \arxivurl#1{\textsf{#1}}\fi
\csname PreBibitemsHook\endcsname

\bibitem{illinois_tool_works_inc_out_2020}
\begin{botherref}
\oauthor{\bsnm{Inc}, \binits{I.T.W.}}:
Out of {Production} 5900 {Series} {Universal} {Testing} {Systems};
(2020).
\url{http://www.instron.com/en/products/testing-systems/universal-testing-systems/low-force-universal-testing-systems/5900-series}
Accessed 2022-06-14
\end{botherref}
\endbibitem

\bibitem{carrera_composite_2016}
\begin{bbook}
\bauthor{\bsnm{Carrera}, \binits{E.}}:
\bbtitle{Composite {Materials} and {Structures} in {Aerospace} {Engineering}:
  {Volume} 828}.
\bpublisher{Trans Tech},
\blocation{Pfaffikon, Switzerland}
(\byear{2016})
\end{bbook}
\endbibitem

\bibitem{simulia_abaqus_2022}
\begin{botherref}
\oauthor{\bsnm{SIMULIA}}:
Abaqus {Unified} {FEA} - {SIMULIA}™ by {Dassault} {Systèmes}®
(2022).
\url{https://www.3ds.com/products-services/simulia/products/abaqus/}
Accessed 2022-06-14
\end{botherref}
\endbibitem

\bibitem{michael_smith_abaqusstandard_2019}
\begin{bbook}
\bauthor{\bsnm{Smith}, \binits{M.}}:
\bbtitle{{ABAQUS}/{Standard} {User}'s {Manual}}.
\bpublisher{Dassault Syst`emes Simulia Corp},
\blocation{United States}
(\byear{2019})
\end{bbook}
\endbibitem

\bibitem{kim_introduction_2008}
\begin{bbook}
\bauthor{\bsnm{Kim}, \binits{N.-H.}},
\bauthor{\bsnm{Sankar}, \binits{B.V.}}:
\bbtitle{Introduction to {Finite} {Element} {Analysis} And {Design}},
\bedition{1st edition} edn.
\bpublisher{Wiley},
\blocation{New York Weinheim}
(\byear{2008})
\end{bbook}
\endbibitem

\bibitem{storakers_material_1986}
\begin{barticle}
\bauthor{\bsnm{Storåkers}, \binits{B.}}:
\batitle{On material representation and constitutive branching in finite
  compressible elasticity}.
\bjtitle{Journal of the Mechanics and Physics of Solids}
\bvolume{34}(\bissue{2}),
\bfpage{125}--\blpage{145}
(\byear{1986}).
\doiurl{10.1016/0022-5096(86)90033-5}.
Accessed 2022-06-14
\end{barticle}
\endbibitem

\end{thebibliography}


\end{document}


\title{Supplementary Material}

\maketitle

\renewcommand*\contentsname{}
\newcommand{\sectionsize}{\fontsize{12}{13}\selectfont}
\tableofcontents

\newpage

\section[Supplementary Figures]{\sectionsize Supplementary Figures}\label{supp:figures}

\begin{figure}[ht]%
\centering
\includegraphics[width=0.75\textwidth]{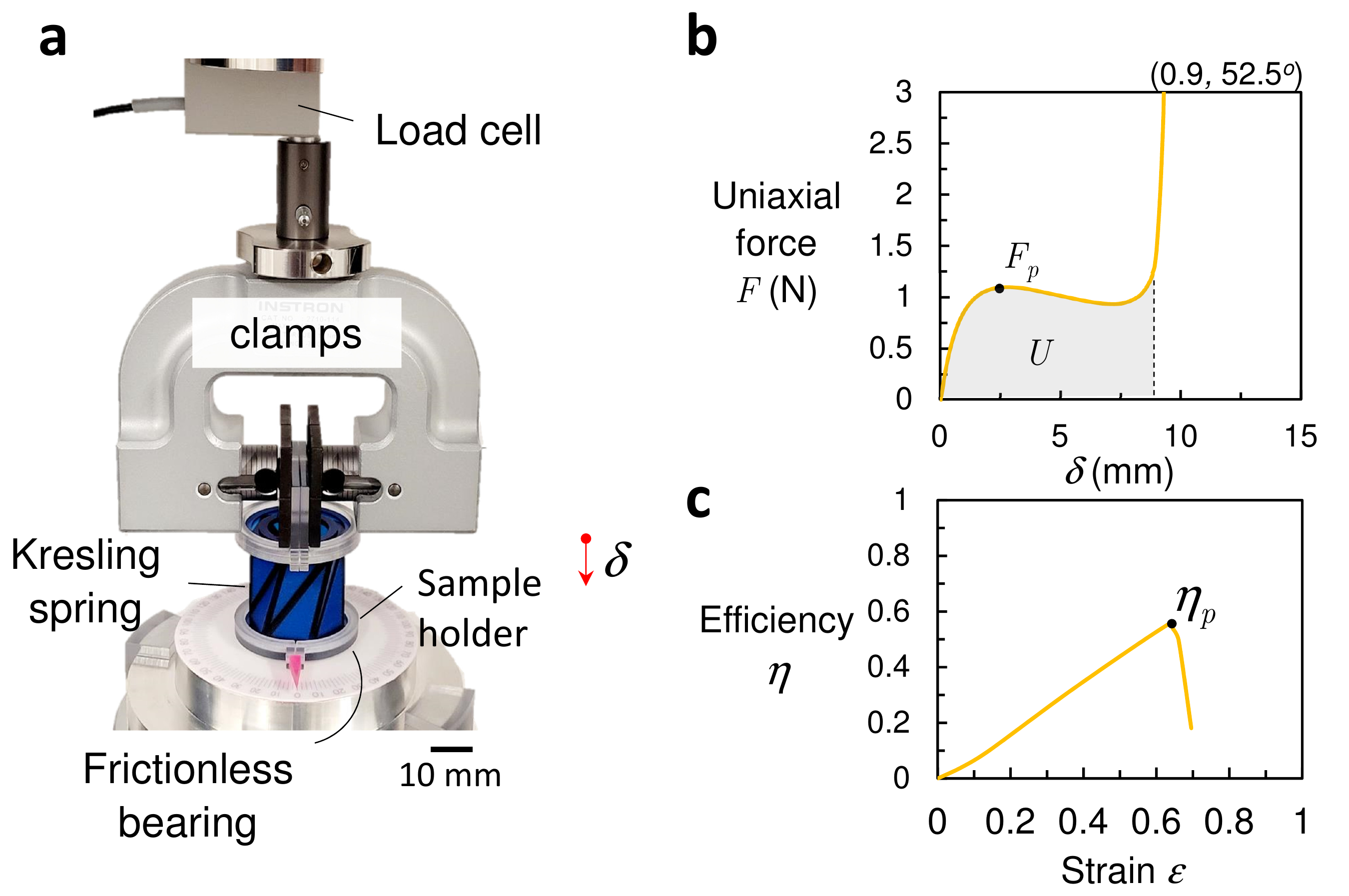}
\caption{ \textbf{Uniaxial testing apparatus and procedure}. \textbf{a} Uniaxial testing rig. \textbf{b} Uniaxial force-deflection curve for ($u_o/R$=0.9, $\phi_o$=52.5$^o$). Shaded grey area represents the total energy absorbed $U$. \textbf{c} Efficiency-strain curve, where peak force and efficiency are denoted by $F_p$ and $\eta_p$ respectively.} \label{fig:S1}
\end{figure}

\begin{figure}[ht]%
\centering
\includegraphics[width=0.75\textwidth]{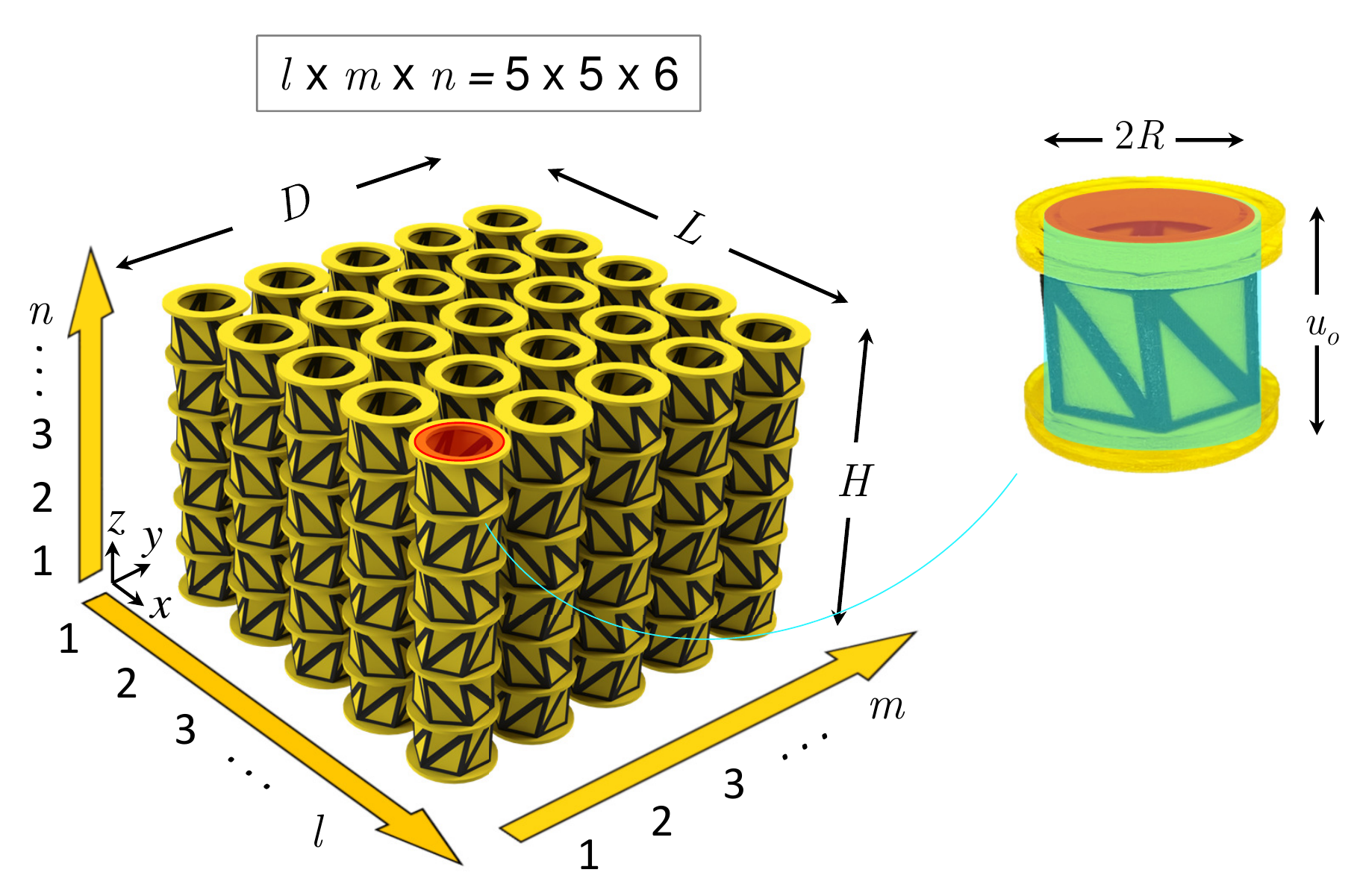}
\caption{ \textbf{Footprint area and volume of unit cells}. The top red colored plane indicates the footprint area of a single unit cell. The volume of the unit cell is indicated by the volume occupied by the cylindrical cyan region. $R$ is the circumscribing radius of the polygon and $u_o$ is the pre-deformation height.} \label{fig:S2}
\end{figure}

\begin{figure}[ht]%
\centering
\includegraphics[width=\textwidth]{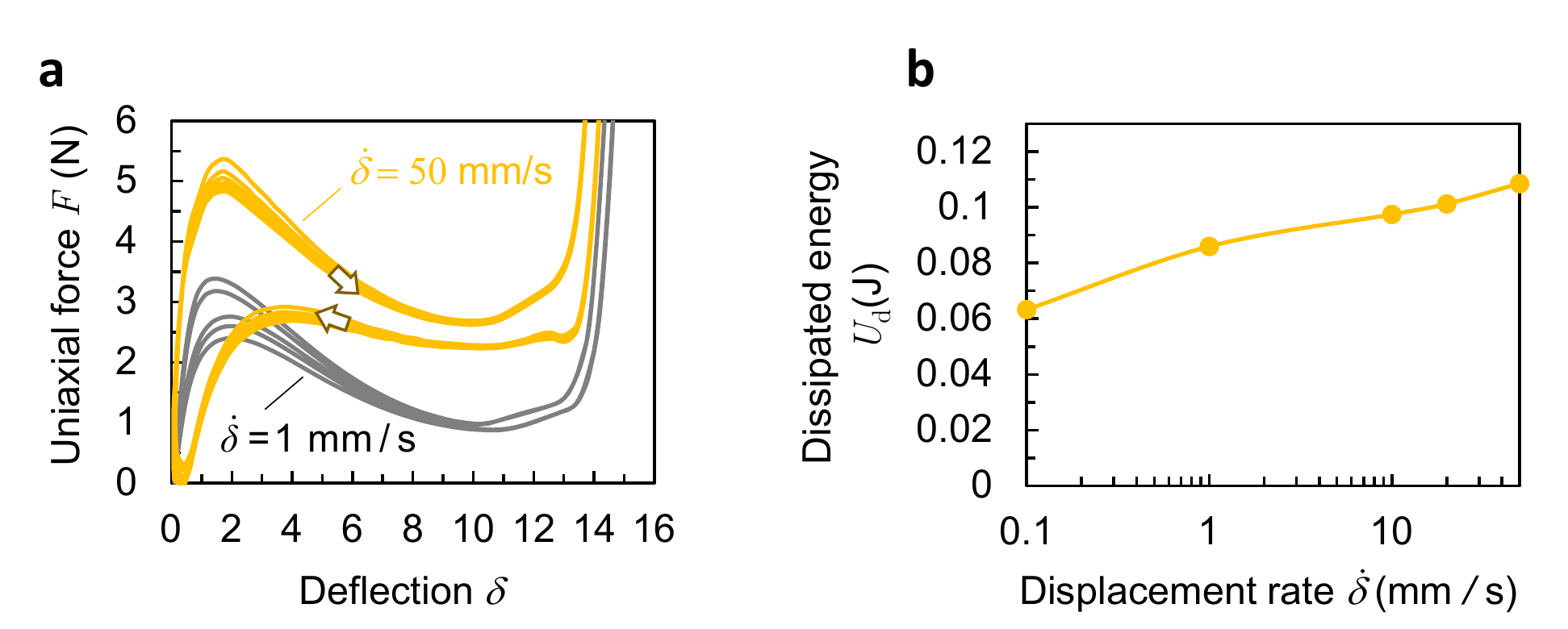}
\caption{ \textbf{Cyclic loading}. \textbf{a} Hysteresis loop obtained under 1 mm/s and 50 mm/s loading rate for 10 cycles. Arrows on curves represent the direction of loading and unloading. \textbf{b} Dissipated energy $U_d$ variation with increased loading rate.} \label{fig:S3}
\end{figure}

\begin{figure}[ht]%
\centering
\includegraphics[width=0.5\textwidth]{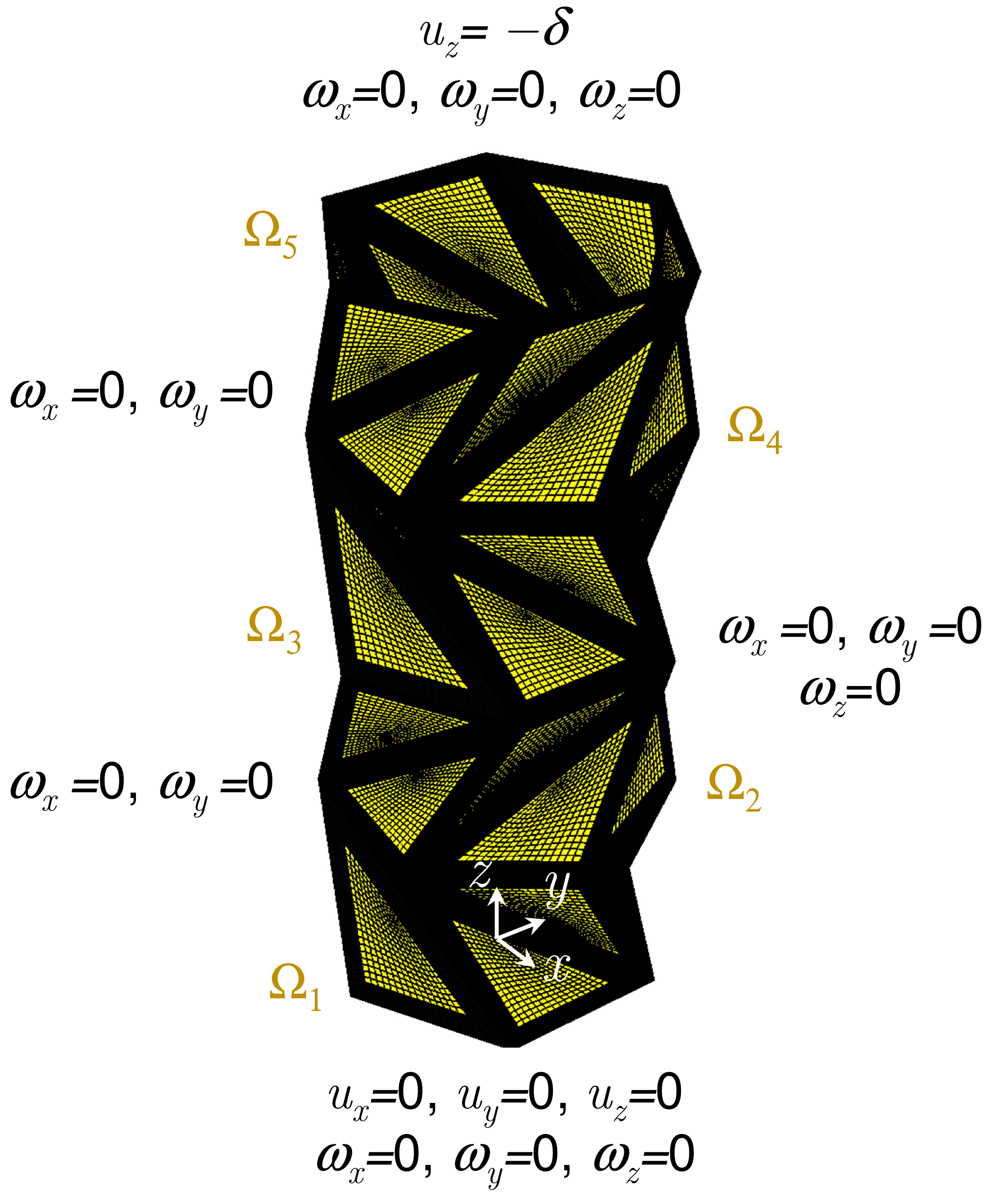}
\caption{ \textbf{Meshed model and applied boundary conditions}. An $l \times m \times n = 1 \times 1 \times 4$ Origami column where the junctions are denoted by $\Omega_i$. The index $i$ indicates the position of junctions. Prescribed displacements along $x$, $y$ and $z$ are denoted by $u_x$, $u_y$ and $u_z$ respectively. Prescribed rotation about $x$, $y$ and $z$ are $\omega_x$, $\omega_y$ and $\omega_z$ respectively.} \label{fig:S4}
\end{figure}

\begin{figure}[ht]%
\centering
\includegraphics[width=\textwidth]{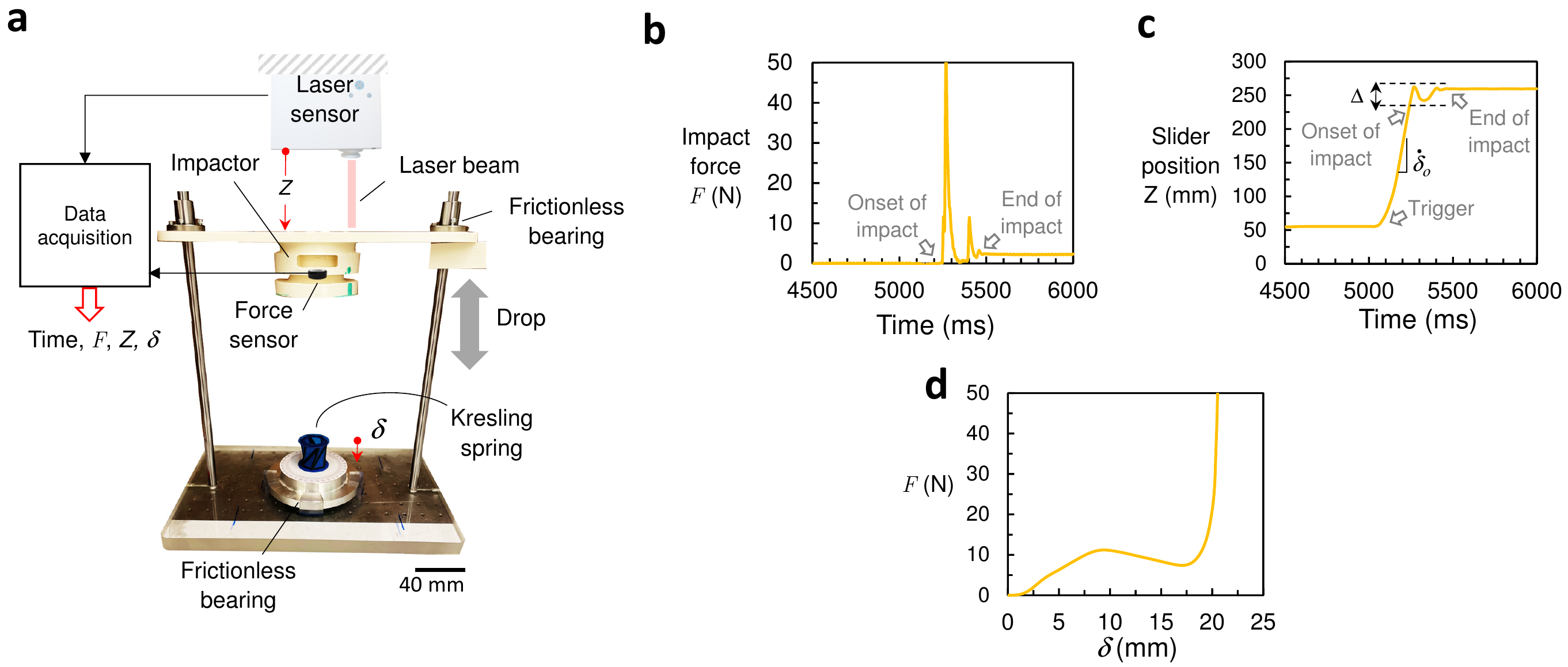}
\caption{ \textbf{Impact testing apparatus and procedure}. \textbf{a} Low-speed impact testing rig that outputs uniaxial impact force $F$, slider position $Z$ and sample deflection $\delta$. \textbf{b} Evolution of $F$ with time during impact. \textbf{c} Slider position evolution with time, where the rebound distance is denoted by $\Delta$, \textbf{d} Impact force-deflection curve. The triggers, onset and end of impact, are indicated by grey arrows.} \label{fig:S5}
\end{figure}

\begin{figure}[ht]%
\centering
\includegraphics[width=\textwidth]{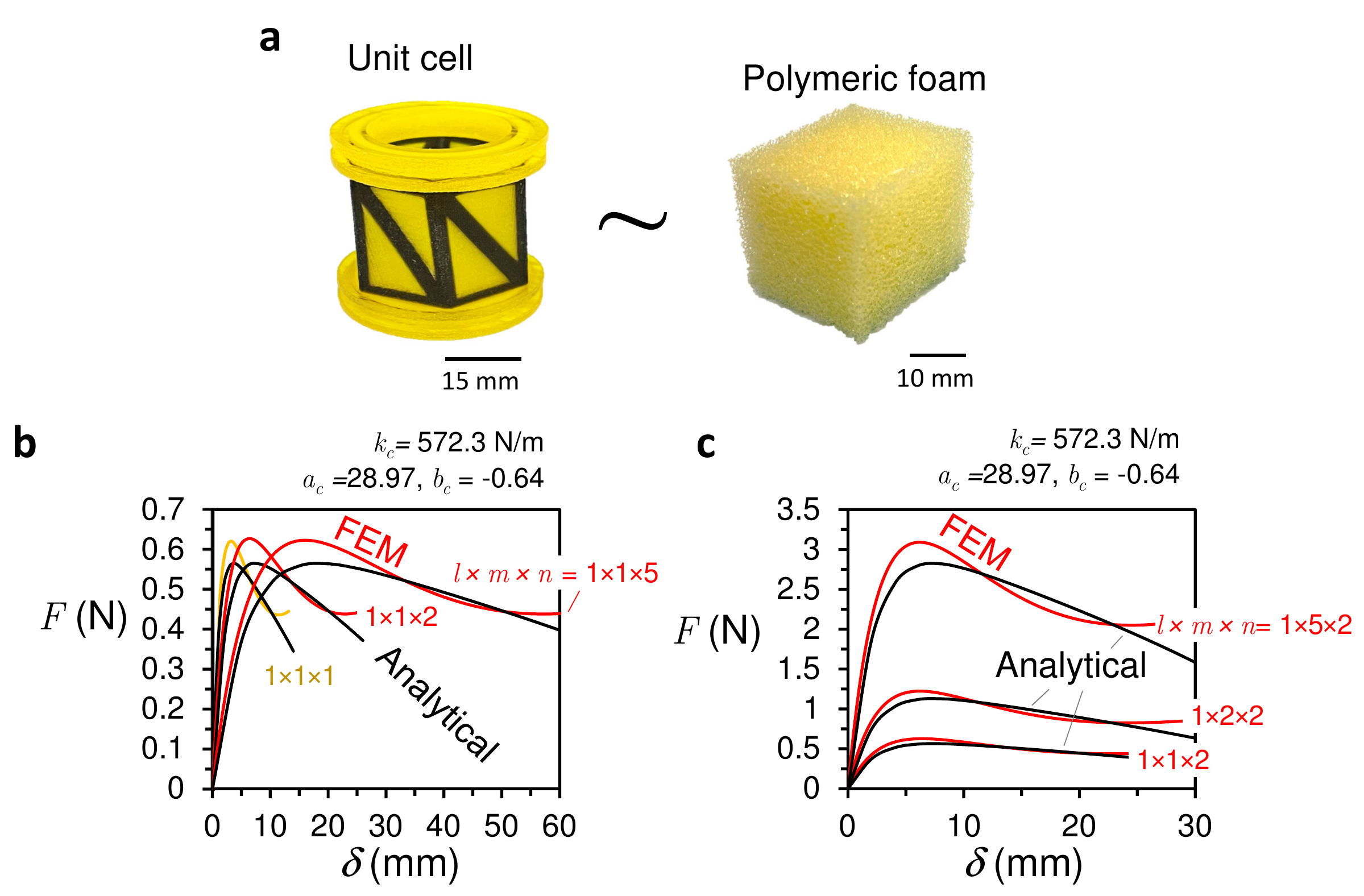}
\caption{ \textbf{Modeling Origami unit cell as a hyperelastic foam}. \textbf{a} Representing the behaviour of the unit cell as polymeric foam. Comparison of the analytical expression with finite element simulations for \textbf{b} series and \textbf{c} planar tessellations.} \label{fig:S6}
\end{figure}

\begin{figure}[ht]%
\centering
\includegraphics[width=\textwidth]{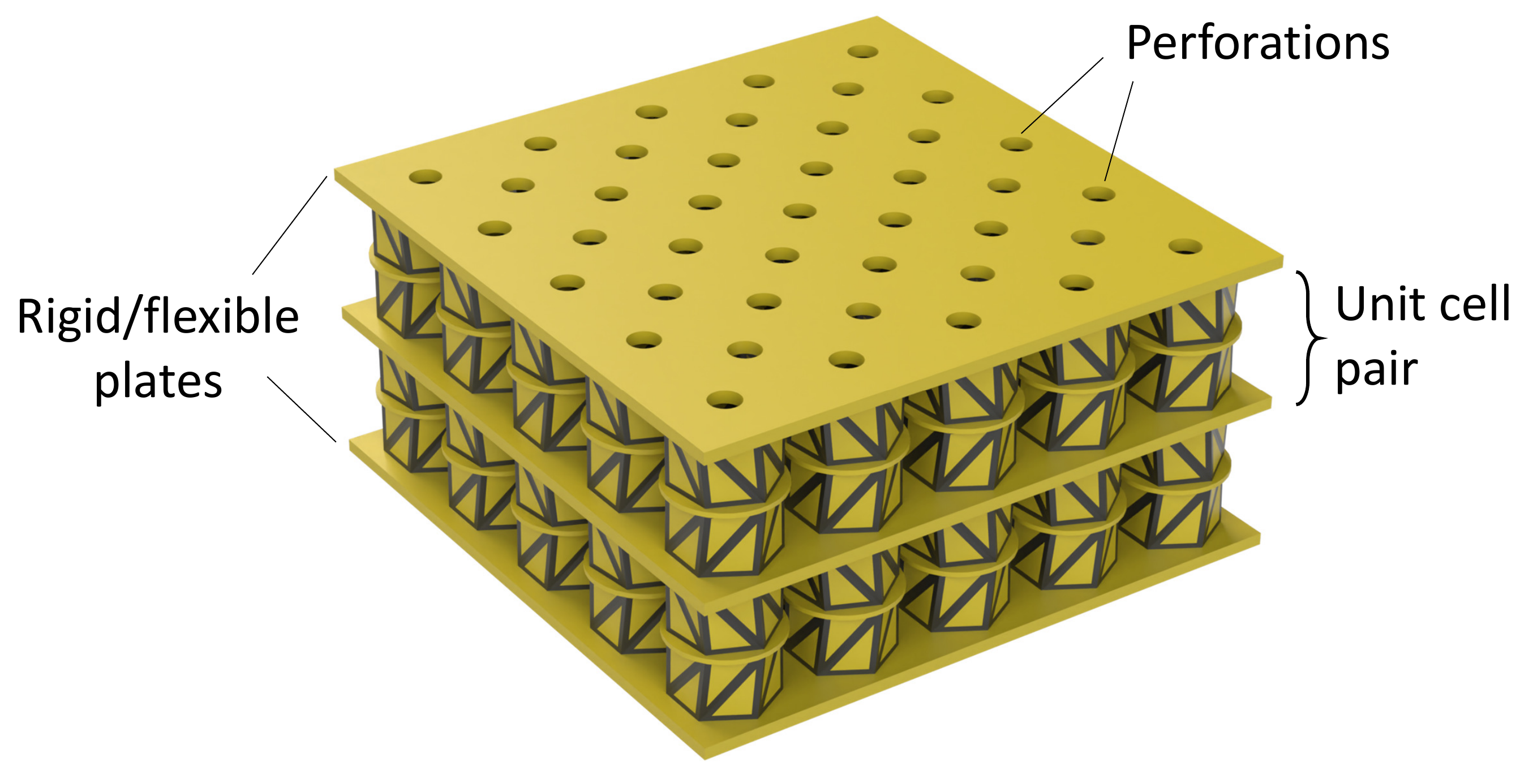}
\caption{ \textbf{Design option}. Tessellated model with perforated top and bottom plates.} \label{fig:S6}
\end{figure}

\clearpage

\section[Supplementary Table]{\sectionsize Supplementary Table}\label{supp:table}

\begin{table}[ht]
\caption{Supplementary table 1: Geometric and material properties. }
\label{tab:my-table}
\begin{tabular}{lllll}
\hline
 \textbf{Geometric Parameters} & & & \textbf{Comment}  \\ \hline
 Circumscribing radius & $R$ & 15 mm & For unit cell (Fig. 3, 4)  \\
 Circumscribing radius & $R$ & 8 mm & For tessellated cushion (Fig. 7, 8)  \\
 Panel thickness & $t$ & 0.75 mm & For unit cell (Fig. 3, 4)\\
 Panel thickness & $t$ & 0.4 mm & For tessellated cushion (Fig. 7, 8) \\
 Flexible frame width (creases) & $w$ & 1.5 mm & For unit cell (Fig. 3, 4)  \\
 Flexible frame width (creases) & $w$ & 0.8 mm & For tessellated cushion (Fig. 7, 8) \\
        & $w/t$ & 2.0 & Fixed for all cases \\
Number of polygon sides & $N$ & 6 & Fixed for all cases \\
\hline
\textbf{Material Parameters} & & &  \\ 
\hline
 Young's modulus of TangoBlackPlus & $E_T$ & 0.3 MPa & \\
 Poisson's ratio of TangoBlackPlus & $\nu_T$ & 0.3 & \\
 Young's modulus of Vero & $E_V$ & 3000 MPa & \\
 Poisson's ratio of Vero & $\nu_V$ & 0.3 & \\

 \hline
\end{tabular}
\end{table}

\clearpage

\section[Supplementary Note 1: Uniaxial Loading Tests]{\sectionsize Supplementary Note 1: Uniaxial Loading Tests}\label{supp:note1}

The uniaxial quasi-static tests were performed using an Instron 5960 series universal testing machine (Instron \cite{illinois_tool_works_inc_out_2020}, Norwood, US). The sample is placed into a sample holder that is designed in-house as shown in Supplementary Fig. \ref{fig:S1}a. The bottom part of the sample holder is connected to a frictionless bearing which permits a smooth rotation of the unit cell upon compression. The entire apparatus rests on a rigid platform. A constant rate displacement is applied at the upper surface of the holder at a rate of 0.15 mm/s up to 80\% of initial height of the tested sample. The resulting axial reaction force $F$, exerted by the sample on the clamp is recorded using a low profile load cell. The measured deflection and reaction force are used to generate the force-deflection ($F-\delta$) curves (Supplementary Fig \ref{fig:S1}b). The resulting $F-\delta$ curves can then be translated into their corresponding engineering stress-strain ($\sigma-\varepsilon$) by: 

\begin{equation}
\sigma=\frac{F}{A}
\end{equation}

\begin{equation}
\varepsilon=\frac{\delta}{u_o}
\end{equation}
where the footprint area (highlighted as red area in Supplementary Fig. \ref{fig:S2}) of a single unit cell is $\pi R^2$. While for a full tessellated structure, the total footprint area is $A=(l m)\pi R^2$. 

The area under the $\sigma-\varepsilon$ curves represent the absorbed energy per volume of the sample, which is also regarded as "toughness" and is given by:
\begin{equation}
Toughness=\int \sigma \text{d}{\varepsilon},
\end{equation} 
The apparent density of the tessellated material can be calculated by:
\begin{equation}
\rho=\frac{M}{V}
\end{equation}
where the volume of single unit cell is $(\pi R^2)u_o$ (indicted by a cyan region in Supplementary Fig. \ref{fig:S2}), while the total volume of  a tessellated material is $V=(lmn)(\pi R^2)u_o$. $M$ is the total mass. The calculated density of our tessellated materials is $\rho = 138.6$ kg/m$^3$.
Let the density of the base constituents, namely Vero and Tango respectively, be $\rho_V$ and $\rho_T$. The effective density of the solid material forming the tessellated material can be expressed using the rule of mixture \cite{carrera_composite_2016} as:
\begin{equation}
\rho_s=f_{_V} \rho_{_V} + f_{_T} \rho_{_T}
\end{equation}
where $f_V$ and  $f_T$ are volume fraction of the Vero and Tango 3D printing materials respectively, that is $\rho_s=1143$ kg/m$^3$. The relative density, which is the apparent density of cellular material to that of the solid material: $\rho/\rho_s$, is 0.12.

Using the above-mentioned uniaxial testing rig, we subjected the unit cell to cyclic displacement driven loading. Supplementary Fig. \ref{fig:S3}a shows the hysteresis for loading rates of 1 mm/s and 50 mm/s. The plot shows also the effect of increased loading rate on inducing viscoelastic hardening where the force is elevated with increased speed. The area enclosed within the hysteresis loop represents the energy dissipated $U_d$ per cycle. This energy dissipation increases with higher loading rate (Supplementary Fig. \ref{fig:S3}b). See also Supplementary Movie 1.

\newpage

\section[Supplementary Note 2: Details of Computational Model]{\sectionsize Supplementary Note 2: Details of Computational Model}\label{supp:note2}

A general shell-based finite element model is built using Abaqus (Dassault Systèmes \cite{simulia_abaqus_2022}, US) to simulate the quasi-static behavior of the Origami-based cellular materials. The material is assumed to behave according to the linear elastic constitutive laws. The model is composed of 3D planar shell elements (S4) that account for 3D translation, rotation, in-plane and out-of-plane deformations \cite{michael_smith_abaqusstandard_2019}. The model takes the geometric parameters $(\frac{u_o}{R}, \phi_o, w/t)$, the material elastic properties ($E_T, \nu_T, E_V, \nu_V)$, and tessellation ($l \times m \times n$) as inputs and uses them to construct the 3D shell-model using the Kresling topology.  Considering an Origami column ($1 \times 1 \times 4$) as an example (Supplementary Fig. \ref{fig:S4}), the bottom surface $\Omega_1$ is fixed, while the top surface $\Omega_{n+1}= \Omega_5$ is subjected to uniaxial compression $u_z =-\delta$. To ensure a well posed boundary value problem with unique solutions \cite{kim_introduction_2008}, all junctions, including top and bottom surfaces ($\Omega_1 - \Omega_{n+1}$), are constrained such that their rotation around the $z$-axis is restricted for odd junctions. The corresponding boundary conditions are:
\begin{align}
&\left\{ \begin{array}{ll}
         u_x=u_y=u_z=0;\\
        \omega_x=\omega_y=\omega_z=0.\end{array} \right. \qquad \text{at}~ \Omega_1,\text{ } z=0\\[5pt]
&\left\{ \begin{array}{ll}
         u_z=-\delta;\\
        \omega_x=\omega_y=\omega_z=0.\end{array} \right. \qquad \text{at}~ \Omega_5,\text{ } z=4 u_o
\end{align}
while the intermediate junctions are constraint by:
\begin{align}
        &\omega_x=\omega_y=0. \qquad \qquad \qquad \text{at}~ \Omega_{2},\text{ } z=u_o \\[5pt]
        &\omega_x=\omega_y=\omega_z=0. \qquad \qquad \text{at}~ \Omega_{3},\text{ } z=2u_o \\[5pt]
        &\omega_x=\omega_y=0. \qquad \qquad \qquad \text{at}~ \Omega_{4},\text{ } z=3u_o
\end{align}
which generalize as:
\begin{align}
    &u_x=u_y=u_z=0. \qquad \text{at}~ \Omega_1,\text{ } z=0 \\[5pt]
    &u_z=-\delta. \qquad \qquad \qquad \quad \text{at}~ \Omega_{n+1},\text{ } z=nu_o \\[5pt]
    &\omega_x=\omega_y=0. \qquad \qquad \text{at}~ \Omega_{i},\text{ } z=(i-1)u_o, \nonumber \\
    &\qquad \qquad \qquad \qquad \qquad i = \text{even integers} \in \{1,2,\ldots, n+1\} \\[5pt]
    &\omega_x=\omega_y=\omega_z=0. \qquad \text{at}~ \Omega_{j},\text{ } z=(j-1)u_o, \nonumber \\
    &\qquad \qquad \qquad \qquad \qquad j = \text{odd integers} \in \{1,2,\ldots, n+1\}
\end{align}
where $u_x, u_y, u_z, \omega_x, \omega_y, \omega_z$ are the displacement and rotation about the $x$, $y$ and $z$ axes, respectively. All junctions are prevented from out-of-plane rotation by enforcing $\omega_x= \omega_y=0$. All remaining surfaces and edges are traction free. 

\newpage

\section[Supplementary Note 3: Analytical Model of \texorpdfstring{$F-\delta$}{Lg} Relationships]{\sectionsize Supplementary Note 3: Analytical Model of $F-\delta$ Relationships}\label{supp:note3}

A single unit cell can be modeled as a hyperelastic compressible material whose strain energy, $W$ can be captured by Storaker’s model \cite{storakers_material_1986} (Supplementary Fig. \ref{fig:S6}a) as
 \begin{equation}
 W=\frac{2\mu}{\alpha^2} \left\{\lambda_1^\alpha+\lambda_2^\alpha+\lambda_3^\alpha-3+\frac{1}{\beta}(J^{-\alpha \beta}-1)\right\},
 \end{equation} 
where $\lambda_1$, $\lambda_2$ and $\lambda_3$ are the principal stretches, which are related to the engineering strain via: $\lambda=\epsilon+1$; $J$ is the Jacobian given by $\lambda_1 \lambda_2\lambda_3$; $\mu$ is the shear modulus, $\alpha$ and $\beta$ are empirical constants used to capturing the nonlinear behavior.

Applying uniaxial compression along the $z$-axis; i.e. along the $\lambda_3$ direction, and assuming that the resulting stretches along the other directions are identical while using the fact that those surfaces are traction free; i.e. $\sigma_1=\sigma_2=0$, we obtain:  
\begin{equation}
 \sigma_3=\frac{\partial W}{\partial \lambda_3} = \frac{F_c}{\pi R^2}=\frac{2 \mu}{\alpha} \left\{\left(1-\frac{\delta}{u_o}\right)^{\alpha-1} -\left(1-\frac{\delta}{u_o}\right)^{-1-\alpha \beta \left(1-\frac{2\beta}{1+2 \beta}\right)}\right\},
 \end{equation} 					  
which can be simplified into the form:
\begin{equation}
 \frac{F_c}{k_c u_o}=\frac{1}{a_c+b_c} \left\{\left(1-\frac{\delta}{u_o}\right)^{a_c} -\left(1-\frac{\delta}{u_o}\right)^{-b_c}\right\},
 \end{equation} 
where $k_c$ is the stiffness of the structure at zero deflection:
\begin{equation}
 k_c=\frac{d F}{d \delta}(\delta=0)=\frac{2\pi \mu R^2}{u_0}\frac{a_c+b_c}{a_c+1},
 \nonumber
 \end{equation} 
and $a_c$ and $b_c$ are empirical constants to be calibrated by high fidelity models or using experiments. 

For unit-cells connected in series, enforcing equilibrium along the $z$-axis of an Origami column requires that the reaction forces balance each other at each junction. Thus, the overall reaction force $F$ must be equal to that at each junction $F_c$; such that
 \begin{equation}
 \begin{split}
F_c=F&=\frac{k_c u_o}{a_c+b_c} \left\{\left(1-\frac{\delta}{u_o}\right)^{a_c} -\left(1-\frac{\delta}{u_o}\right)^{-b_c}\right\}\\
&= \frac{k H}{a+b} \left\{\left(1-\frac{\delta}{u_o}\right)^{a} -\left(1-\frac{\delta}{u_o}\right)^{-b}\right\}.
\end{split}
\label{FcEqn}
 \end{equation} 
This yields $a_c=a$, $b_c=b$, and $k=k_c/n$.
						    
For unit cells connected in parallel; i.e., in-plane tessellation ($l \times m \times 1$), invoking equilibrium along $z$ axis yields:
\begin{equation}
F=\sum_{c=1}^{lm} F_c.
\label{sumFc}
\end{equation}
This results in $a=a_c$, $b=b_c$ and $k=(lm)k_c$. Combining Eq. (\ref{FcEqn}) with Eq. (\ref{sumFc}) yields:
\begin{equation}
\frac{F}{k H}=\frac{lm}{n}\frac{1}{a_c+b_c} \left\{\left(1-\frac{\delta}{H}\right)^{a_c} -\left(1-\frac{\delta}{H}\right)^{-b_c}\right\},
\label{finalF}
\end{equation}
Equation (\ref{finalF}) is then used to obtain an analytical expression for the $F-\delta$ curve of the Origami-based cellular material. First, we calibrate the constants $k_c$, $a_c$, and $b_c$ at the level of the unit cell to match the experimental data. Second, those constants can then be used to predict the overall response for an any arbitrary tessellation (\textit{l}×\textit{m}×\textit{n}) using Eq. (\ref{finalF}). For example, Supplementary Fig \ref{fig:S6}b shows how the analytical form (in solid black line) fits the experimental measurements for a single unit cell (denoted as a solid yellow line) for $k_c=572.3$ N/m, $a_c=28.97$ and $b_c=-0.64$. Finally, using those calibrated constants, we predicted the profile for different sets of \textit{l}×\textit{m}×\textit{n} cases. Both numerical simulations (in solid red lines) with their corresponding analytical predictions (in solid black lines) are compared on Supplementary Fig. \ref{fig:S6}b and c.

Since the area under the $F-\delta$ curve provides the total energy absorbed, 
\begin{equation}
U=\int F \text{d}{\delta},
\end{equation} 
the total energy absorbed by the cushion can be written as:
\begin{equation}
U=(l m n) \int F_c \text{d}{\delta},
\label{scaledU}
\end{equation}

\newpage

\section[Supplementary Movies]{\sectionsize Supplementary Movies}\label{supp:movies}

\textbf{Filename:} Supplementary Movie 1 - Hysteresis Loop.mp4 \\
\textbf{Description:} Cyclic loading performed on a single unit cell at 20 mm/s. This video shows the generated hysteresis effect and the energy dissipation capacity of the unit cell. 
\\[10pt]
\textbf{Filename:} Supplementary Movie 2 - Impact Video.mp4 \\
\textbf{Description:} High speed imaging showing free fall impact test performed on the tessellated cushion ($l \times m \times n = 5 \times 5 \times 4$) at an approaching speed of 1.85 m/s. The video is captured at 10,000 frames/seconds (fps). 
\\[10pt]
\textbf{Filename:} Supplementary Movie 3 - Recovery.mp4 \\
\textbf{Description:} Demonstrating the shape recovery of cushion material by bare hands compression.

\newpage

\bibliography{References}
